\documentclass[journal]{IEEEtran}
\IEEEoverridecommandlockouts
\usepackage{cite}

\usepackage{amsmath,amssymb,amsfonts,mathtools,bm}
\usepackage{amsthm}
\usepackage{algorithm}
\usepackage{algpseudocode}
\usepackage{graphicx}
\usepackage[caption=false,font=footnotesize]{subfig}
\usepackage{textcomp}
\usepackage{xcolor,float}
\usepackage{tabularx}
\usepackage{textcomp}
\usepackage{diagbox}
\usepackage{svg}
\usepackage{booktabs}
\usepackage{hyperref}
\hypersetup{hidelinks}
\newcommand{\added}[1]{#1}

\usepackage{booktabs}
\usepackage{array} 

\usepackage{multirow}

\usepackage{tikz,xcolor,hyperref}
\definecolor{lime}{HTML}{A6CE39}
\DeclareRobustCommand{\orcidicon}{%
    \begin{tikzpicture}
    \draw[lime, fill=lime] (0,0) 
    circle [radius=0.16] 
    node[white] {{\fontfamily{qag}\selectfont \tiny ID}};    \draw[white, fill=white] (-0.0625,0.095) 
    circle [radius=0.007];    \end{tikzpicture}
    \hspace{-2mm}}
\foreach \x in {A, ..., Z}{%
    \expandafter\xdef\csname orcid\x\endcsname{\noexpand\href{https://orcid.org/\csname orcidauthor\x\endcsname}{\noexpand\orcidicon}}
    }

\allowdisplaybreaks
\renewcommand{\baselinestretch}{1}

\begin{document}

\title{PrismWF: A Multi-Granularity Patch-Based Transformer for Robust Website Fingerprinting Attack}

    

\author{
Yuhao Pan\orcidA{},
Wenchao Xu\orcidB{},~\IEEEmembership{Member,~IEEE},
Fushuo Huo\orcidC{},
Haozhao Wang\orcidD{},~\IEEEmembership{Member,~IEEE},
Xiucheng Wang\orcidE{},~\IEEEmembership{Graduate Student Member,~IEEE,}
Nan Cheng\orcidF{},~\IEEEmembership{Senior Member,~IEEE}

\thanks{ }
\thanks{
\par Yuhao Pan and Wenchao Xu are with the Division of Integrative Systems and Design, Hong Kong University of Science and Technology, Hong Kong, China (e-mail: ypanca@connect.ust.hk, wenchaoxu@ust.hk). \textit{Wenchao Xu is the corresponding author}.

\par Xiucheng Wang and Nan Cheng are with the State Key Laboratory of ISN and School of Telecommunications Engineering, Xidian University, Xi’an 710071, China (e-mail: xcwang\_1@stu.xidian.edu.cn, dr.nan.cheng@ieee.org). 

\par Fushuo Huo is with the School of Cyber Science and Engineering, Southeast University, Nanjing, China (e-mail: fushuohuo@seu.edu.cn).

\par Haozhao Wang is with the School of Computer Science and Technology, Huazhong University of Science and Technology, Wuhan, 430068, China (email: hz\_wang@hust.edu.cn).
}
}

    \maketitle

\IEEEdisplaynontitleabstractindextext

\IEEEpeerreviewmaketitle

\begin{abstract}

Tor is a low-latency anonymous communication network that protects user privacy by encrypting website traffic. However, recent website fingerprinting (WF) attacks have shown that encrypted traffic can still leak users' visited websites by exploiting statistical features such as packet size, direction, and inter-arrival time. Most existing WF attacks formulate the problem as a single-tab classification task, limiting their effectiveness when concurrent website visits produce mixed traffic traces. To this end, we propose PrismWF, a multi-granularity patch-based Transformer for multi-tab WF attacks. Specifically, we design a robust traffic feature representation for raw web traffic traces and extract multi-granularity features using convolutional kernels with different receptive fields. To effectively integrate information across temporal scales, the proposed model refines features through three hierarchical interaction mechanisms: inter-granularity detail supplementation from fine to coarse granularities, intra-granularity patch interaction with dedicated router tokens, and router-guided dual-level intra- and cross-granularity fusion. This design aligns with the cognitive logic of global coarse-grained reconnaissance and local fine-grained querying, enabling effective modeling of mixed traffic patterns in WF attack scenarios. Extensive experiments on public benchmarks, WF defenses, and real-world Tor deployments demonstrate that our method achieves state-of-the-art performance compared to existing baselines. \added{Code is available at \url{https://github.com/yyyyu120/PrismWF}.}

\end{abstract}


\section{Introduction}
In recent years, the rapid growth of the Internet has intensified concerns over user privacy during web browsing.
To mitigate these risks, anonymous communication systems have been widely deployed to conceal users’ identities and browsing behaviors.
Among them, the Tor network is one of the most widely used low-latency anonymity systems, serving millions of users worldwide~\cite{dingledine2004tor, mani2018understanding}.
Tor achieves anonymity by routing encrypted traffic through a multi-hop overlay network of relay nodes, such that no single relay can associate a user with their destination.

\added{Despite these protections, the encrypted traffic between the Tor client and the entry node remains observable to a local adversary.}
By analyzing observable traffic characteristics, such as packet sizes, packet directions, and inter-packet timing patterns, a local adversary can still infer sensitive information about users’ web activities.
Early website fingerprinting (WF) attacks primarily relied on expert knowledge to manually design discriminative traffic features, which were then fed into traditional machine learning classifiers (e.g., random forests, support vector machines, and \emph{k}-nearest neighbors) for website identification \cite{wang2014effective, panchenko2016website, hayes2016k}.
With the rapid advancement of deep learning (DL) \cite{he2016deep}, subsequent studies have proposed end-to-end neural network–based models that automatically learn discriminative representations directly from raw traffic traces, significantly improving attack performance \cite{rimmer2017automated, bhat2018var, sirinam2019triplet}.
Moreover, as various mechanisms for defense against WF have been proposed, robust attack models (e.g., Deep Fingerprinting~\cite{sirinam2018deep} and Robust Fingerprinting~\cite{shen2023subverting}) have been developed to maintain effectiveness under defended traffic conditions.
These models are designed to mitigate performance degradation caused by traffic obfuscation and defense strategies, enabling more resilient WF attacks.

Despite their effectiveness, existing single-tab WF attack methods still suffer from notable limitations.
Most prior approaches are developed under the core assumption that users access only one website at a time, where traffic from different domains is not interleaved.
However, in realistic browsing scenarios, users frequently visit multiple websites concurrently, resulting in mixed traffic traces with interleaved packets from different websites.
As a result, single-tab attack models are inherently limited and often struggle to achieve reliable performance under such traffic mixing conditions.
To address this gap, a number of multi-tab WF attack methods have been proposed to identify multiple coexisting websites from a single mixed traffic trace.
Nevertheless, existing multi-tab approaches still exhibit critical limitations that hinder their practical deployment and effectiveness.
On the one hand, some methods lack robust traffic representation design and directly leverage raw traffic traces as input, which renders them incapable of accurately modeling the complex interleaving patterns of multi-tab traffic \cite{guan2021bapm, jin2023transformer}.
On the other hand, other approaches adopt general DL architectures without adequately considering the mixed traffic traces of multi-tab browsing, leading to suboptimal performance.
Furthermore, a subset of multi-tab WF attacks requires \emph{a priori} knowledge of the exact number of concurrent websites present in the traffic trace, which is an unrealistic assumption in practical attack scenarios, undermining the methods’ real-world applicability \cite{xu2018multi, guan2021bapm}.

Motivated by these limitations, we propose \textbf{PrismWF}, a multi-granularity patch-based transformer tailored to the mixed characteristics of unidentified WF traffic, which explicitly models cross-temporal information interactions for robust multi-tab WF attacks.
Concretely, we first transform raw website traffic traces into a more robust representation that captures essential traffic characteristics.
Based on this representation, multi-granularity traffic features are extracted via parallel convolutional branches, yielding both coarse-grained global features and fine-grained local features.
To model the intrinsic information flow of website traffic, we design a traffic-aware interaction mechanism that enables structured information exchange across granularities.
Coarse-grained features capture broader contextual patterns of traffic segments, while fine-grained features preserve detailed temporal variations, allowing complementary information to be effectively integrated across scales.
In addition, we introduce granularity-specific routers to aggregate semantic information within each granularity, and concatenate router representations from all granularities for final website identification.
This design substantially mitigates information loss caused by traffic mixing in multi-tab browsing scenarios, leading to improved robustness and accuracy in WF attacks.

Extensive experiments demonstrate that PrismWF consistently achieves state-of-the-art performance under multi-tab WF settings.
\added{Moreover, PrismWF remains robust under various WF defenses, maintains stable performance as the number of concurrently opened tabs increases, and generalizes across different tab counts after unified mixed-tab training.}
The main contributions of this paper are summarized as follows:
\begin{enumerate}
    \item 
    We propose PrismWF, a robust multi-tab WF attack model explicitly designed to address traffic mixing caused by concurrent multi-tab browsing.
    \item 
    We introduce a novel transformer-based architecture built upon Multi-Granularity Attention Blocks, which jointly model inter-granularity and intra-granularity traffic patterns across fine-to-coarse temporal scales.
    A dedicated router mechanism is further introduced to aggregate cross-granularity traffic cues, effectively mitigating performance degradation caused by traffic mixing in multi-tab scenarios.
    \item 
    \added{We evaluate PrismWF on public benchmarks covering closed-world, open-world, mixed-tab, and representative simulated-defense settings, as well as in real-world Tor deployments under undefended and online defended conditions. We further analyze statistical significance and computational efficiency.}
\end{enumerate}

The remainder of the paper is organized as follows. We first introduce the related work in Section \ref{sec2} and the threat model in Section \ref{sec3}. We present concrete design of PrismWF in Section \ref{sec4} from robust trace representation, website feature extraction, multi-granularity attention block, and website identification. Next, we conduct a comprehensive evaluation on the performance of PrismWF in Section \ref{sec5}. We discuss relevant issues in Section \ref{sec6} and conclude this paper in Section \ref{sec_conclusion}.

\section{Related Work}
\label{sec2}
\subsection{WF Attack}

\noindent\textbf{Single-tab WF attacks.} 
Early WF attacks primarily relied on expert prior knowledge to transform raw traffic traces into hand-crafted features, followed by traditional machine learning techniques for website identification \cite{panchenko2011website, cai2012touching, panchenko2016website, wang2013improved}.
Wang et al.~\cite{wang2014effective} extracted a large set of statistical features and improved \emph{k}-NN classification through a weighted distance metric.
Panchenko et al.~\cite{panchenko2016website} proposed CUMUL, which uses cumulative packet size features to represent traffic traces and employs an SVM classifier.
To improve robustness against noise and traffic perturbations, Hayes et al.~\cite{hayes2016k} introduced the \emph{k}-fingerprinting approach, which maps hand-crafted features into fingerprint representations via random forests and performs website identification using \emph{k}-NN.

With the strong capability of DL models for end-to-end feature learning, Rimmer et al.~\cite{rimmer2017automated} first introduced the AWF framework, enabling WF attacks without manual feature engineering. 
Building upon this direction, Sirinam et al. \cite{sirinam2018deep} proposed DF, which adopts a deeper and more expressive CNN architecture. The DF model achieves nearly 98\% classification accuracy in the closed-world setting and maintains around 90\% accuracy even under the WTF-PAD defense.
Bhat et al. \cite{bhat2018var} presented Var-CNN, a dual-branch residual convolutional network. The two branches independently model packet direction sequences and temporal traffic variations, and their features are fused to improve WF performance, particularly under limited training data.
To further enhance portability and practical applicability, Sirinam et al. \cite{sirinam2019triplet} explored an \emph{n}-shot learning framework based on triplet loss. By performing metric learning in the feature space, this approach pulls traffic traces from the same website closer together while pushing those from different websites farther apart, enabling more flexible and transferable WF.
Recently, Shen et al. \cite{shen2023subverting} proposed the Traffic Aggregation Matrix (TAM) traffic representation, which aggregates uplink and downlink packets within fixed time slots to construct a robust traffic representation, enabling robust WF attacks. 
Deng et al. \cite{deng2024robust} proposed an early-stage WF attack based on spatio-temporal distribution analysis. By aligning features extracted from early traffic and adopting supervised contrastive learning, their method enables accurate website identification at the early stage of page loading.
\added{Recent WF studies also consider variations in network conditions and observation positions. Swallow~\cite{shen2025swallow} introduces consistent interaction features to support rapid transfer to new network conditions while maintaining robustness against WF defenses, whereas CellShift~\cite{jansen2026cellshift} uses RTT-aware trace transduction to transform exit-side traces into approximations of entry-side observations. Overall, prior single-tab studies address traffic representation, early inference, cross-condition transfer, and observation-position adaptation, while PrismWF focuses on multi-tab identification by modeling concurrently mixed traffic across temporal granularities.}

\begin{figure}[!t]
    \centering
    \includegraphics[width=0.9\columnwidth]{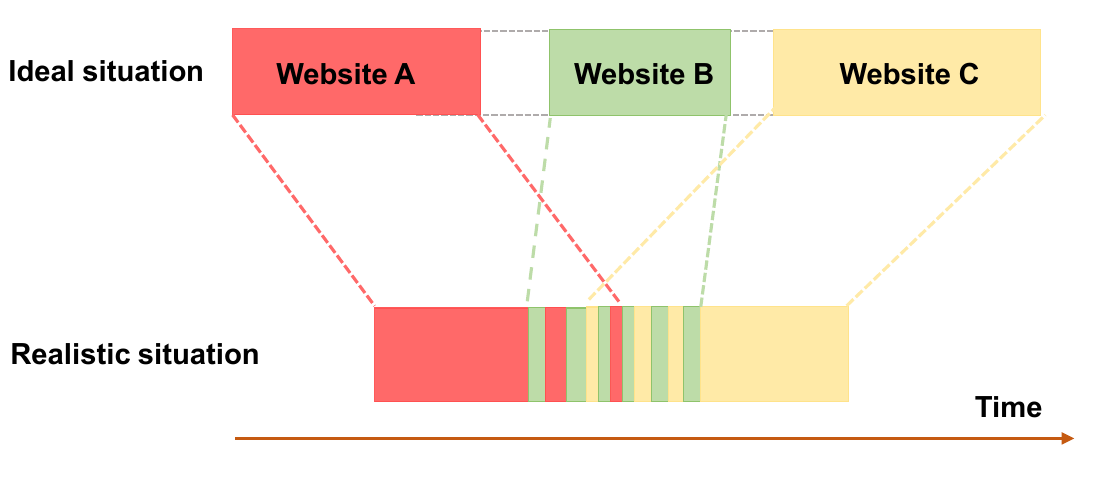}
    \caption{Illustration of traffic mixing caused by concurrent multi-tab browsing. \added{In the idealized sequential case, website-specific traffic segments remain separable, whereas concurrent page loads interleave traffic from multiple websites within the observed trace.}}
    \label{Tor}
\end{figure}

\noindent\textbf{Multi-tab WF attacks.} 
Most existing WF methods rely on a single-tab assumption, where users are presumed to access only one website at a time.
This assumption does not hold in practice, as multi-tab browsing generates mixed traffic traces that substantially degrade the effectiveness of single-tab models.
To bridge this gap, researchers have started exploring multi-tab WF attacks, which aim to identify multiple website labels from a single mixed traffic trace.

In the field of multi-tab WF attacks, Guan et al.~\cite{guan2021bapm} proposed BAPM, one of the first deep-learning-based end-to-end multi-tab WF attack models. BAPM performs block-level modeling on mixed traffic traces to learn tab-aware representations and employs a self-attention mechanism to adaptively fuse information across traffic blocks.
Jin et al. \cite{jin2023transformer} introduced a Transformer-based WF attack model inspired by the DETR-style encoder--decoder architecture \cite{carion2020end}. Their method adopts the feature extraction module from DF \cite{sirinam2018deep} as a backbone to preprocess raw traffic traces into feature sequences. A Transformer encoder is then used to capture global traffic context, while tab-aware query vectors in the decoder interact with encoded features to enable multi-tab website identification.
Deng et al. \cite{deng2023robust, deng2026towards} proposed ARES, which first transforms multi-tab website traffic into the MTAF feature representation. 
Subsequently, multi-head attention is employed to capture global traffic dependencies, together with a Top-$K$ sparse attention mechanism that focuses on traffic patterns most relevant to target labels. 
More recently, Deng et al.~\cite{deng2025countmamba} proposed CountMamba, which employs a causal CNN to extract traffic features and leverages a Mamba state space model for multi-tab website fingerprinting attacks.
\added{Recent multi-tab studies have also examined data efficiency and adaptive segmentation. Meng et al.~\cite{meng2025fmwf} proposed FMWF, which synthesizes realistic multi-tab traffic sequences from single-tab traces and adapts a pretrained model to new multi-tab environments with limited additional data. Guo and Chen~\cite{guo2026adaptive} use label-aware matching scores to adjust sliding-window positions and lengths, fuse segment features through label-aware attention, and incorporate incremental learning for evolving website categories. Together with earlier multi-tab attacks, these recent methods illustrate a broader progression from end-to-end classification toward specialized designs for traffic representation, sequence modeling, data-efficient adaptation, and adaptive segmentation. Within this broader line of research, PrismWF focuses on structured interactions across multiple temporal granularities in directly collected concurrent traces.}

\subsection{WF Defense}
To defend against WF attacks in Tor, researchers have proposed to obfuscate traffic traces by delaying real packets and injecting dummy packets, thereby effectively degrading the performance of WF attack models \cite{mathews2023sok, cui2025comprehensive}.
\added{Existing defenses can be broadly categorized into regularization-based and adversarial example (AE)-based WF defenses.}

\noindent\textbf{Regularization-based WF Defenses.} 
Early WF defenses obfuscate traffic fingerprints by enforcing fixed-time-interval transmissions with uniform packet sizes.
BuFLO \cite{dyer2012peek} adopts a strict constant sending interval, pads all packets to a uniform size, and extends transmissions to a predefined maximum duration—effectively eliminating timing and size-based discriminative features.
Tamaraw \cite{cai2014systematic} improves upon BuFLO by supporting asymmetric constant rates for upstream and downstream traffic, which better matches the inherent asymmetry of web communications and reduces overhead.
Despite this optimization, both approaches incur non-trivial bandwidth and latency overhead, severely limiting their practical deployment in real-world scenarios.
To improve deployability, Juárez et al. \cite{juarez2016toward} propose WTF-PAD, an adaptive padding defense that leverages statistical models of traffic bursts to obfuscate website fingerprints without enforcing constant-rate transmission.
\added{Gong et al. \cite{gong2020zero} propose FRONT, a zero-delay and lightweight WF defense that injects dummy packets during idle periods.}
The padding intervals are randomly sampled from a Rayleigh distribution to obfuscate timing patterns.

\noindent\textbf{AE-based WF Defenses.} 
Recently, AE-based techniques have been integrated into WF defenses, by injecting dummy packets to balance defensive efficacy and bandwidth overhead without additional latency. Mockingbird \cite{rahman2020mockingbird} employs a targeted strategy that iteratively perturbs traffic traces toward a selected target to reduce distinguishability. Sadeghzadeh et al. \cite{sadeghzadeh2021awa} introduce AWA, constructing perturbations based on pairwise website relationships with two variants: NUAWA, which derives sample-specific perturbations, and UAWA, which learns a universal noise-driven perturbation for cross-website reuse. 
More recently, ALERT \cite{qiao2024trace} utilizes a generator to produce targeted adversarial traffic, not only weakening adversarially trained attack models but also achieving state-of-the-art defense performance. 
However, as these methods often suffer from degraded effectiveness in model transfer scenarios, we focus on evaluating WF attacks under regularization-based defense settings.

\subsection{WF Trace Representation}
WF attacks identify target websites solely through traffic analysis based on packet-level metadata, such as packet direction and timestamps.
Early WF attack methods, including AWF~\cite{rimmer2017automated} and DF~\cite{sirinam2018deep}, primarily relied on packet direction sequences (i.e., $+1$ for outgoing packets and $-1$ for incoming packets).
Subsequent work, such as Tik-Tok~\cite{rahman2019tik}, incorporated temporal information by jointly modeling packet directions and timestamps, leading to improved classification performance.
Building on this line of research, Shen et al.~\cite{shen2023subverting} argued that transforming raw traffic traces into coarse-grained temporal representations can yield more robust attack performance, and proposed the TAM representation— which segments traffic into fixed-length time windows to form a $2 \times N$ matrix ($N = \lceil T / d \rceil$, with $T$ as maximum page loading time and $d$ as window size)—for recording incoming and outgoing packet counts.
Following this paradigm, a variety of window-based and statistical representations have been explored.
For example, LASERBEAK~\cite{mathews2024laserbeak} employed multi-dimensional statistical feature vectors,
WFCAT~\cite{gong2025wfcat} captured temporal characteristics using logarithmically binned inter-arrival time (IAT) histograms,
and ARES~\cite{deng2026towards} proposed the MTAF representation, which extracts window-level features to jointly model cell-level and burst-level traffic patterns.

\section{Threat model}
\begin{figure}[ht]
    \centering
    \includegraphics[width=0.9\columnwidth]{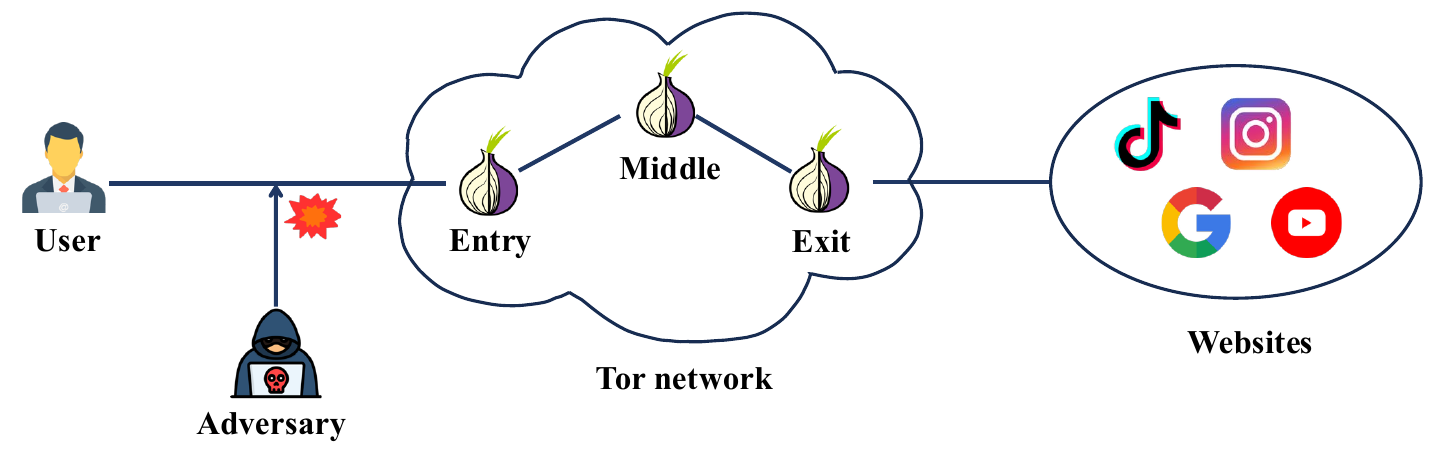}
    \caption{\added{Threat model for website fingerprinting in Tor. A passive local adversary observes encrypted traffic between the Tor client and entry relay and infers concurrently visited websites from packet timing, size, and direction.}}
    \label{Tor1}
\end{figure}
\label{sec3}

As illustrated in Fig.~\ref{Tor1}, users achieve anonymous web browsing via the Tor network, where traffic is routed through a circuit of three randomly selected relay nodes—this multi-hop mechanism prevents destination websites from directly identifying users during access. 
In the WF threat model, the adversary is assumed to be positioned between the user and the Tor entry node.
Due to end-to-end encryption, the adversary cannot decrypt packet payloads and can only leverage traffic metadata (e.g., packet timing, size, and direction) to infer the user’s visited websites, formulating WF attacks as a traffic classification task. Following prior WF studies \cite{sirinam2018deep, sirinam2019triplet, deng2025countmamba, jin2023transformer}, the adversary is restricted to passive observation, with no ability to modify, inject, or drop network packets.

In a typical WF attack pipeline, the adversary first collects encrypted traffic traces by visiting a set of monitored websites via the Tor network to construct a labeled training dataset. 
A WF attack model is then trained to associate encrypted traffic patterns with website identities. 
During deployment, the adversary passively observes the target user’s encrypted traffic and uses the trained model to infer visited websites.

This study considers a more realistic and challenging WF attack scenario where users access multiple websites concurrently. 
Such concurrent access causes traffic mixing in aggregated traces, significantly degrading traditional single-tab WF attack performance. 
We further consider a practical mixed-tab setting, where the adversary has no prior knowledge of the number of concurrently accessed websites in collected traces. 
We also account for WF defense mechanisms: given the high cost of customized defenses, users typically rely on publicly available alternatives in practice.

WF attacks are commonly evaluated under two standard settings~\cite{sirinam2018deep, deng2023robust, gong2025wfcat, deng2025countmamba, rimmer2017automated}: the closed-world and open-world scenarios.
Given the vast number of websites on the Internet, attackers typically monitor only a subset of websites of specific interest.
In the closed-world scenario, the attack task is formulated as classifying traffic traces exclusively into the predefined set of monitored websites.
In the more realistic open-world scenario, users may also visit a large number of websites outside the monitored set, referred to as unmonitored websites.
To distinguish between monitored and unmonitored traffic, prior work typically aggregates all unmonitored websites into an additional class, constructing a practical open-world evaluation setting.

\begin{figure*}[ht]
    \centering
    \includegraphics[width=0.96\textwidth]{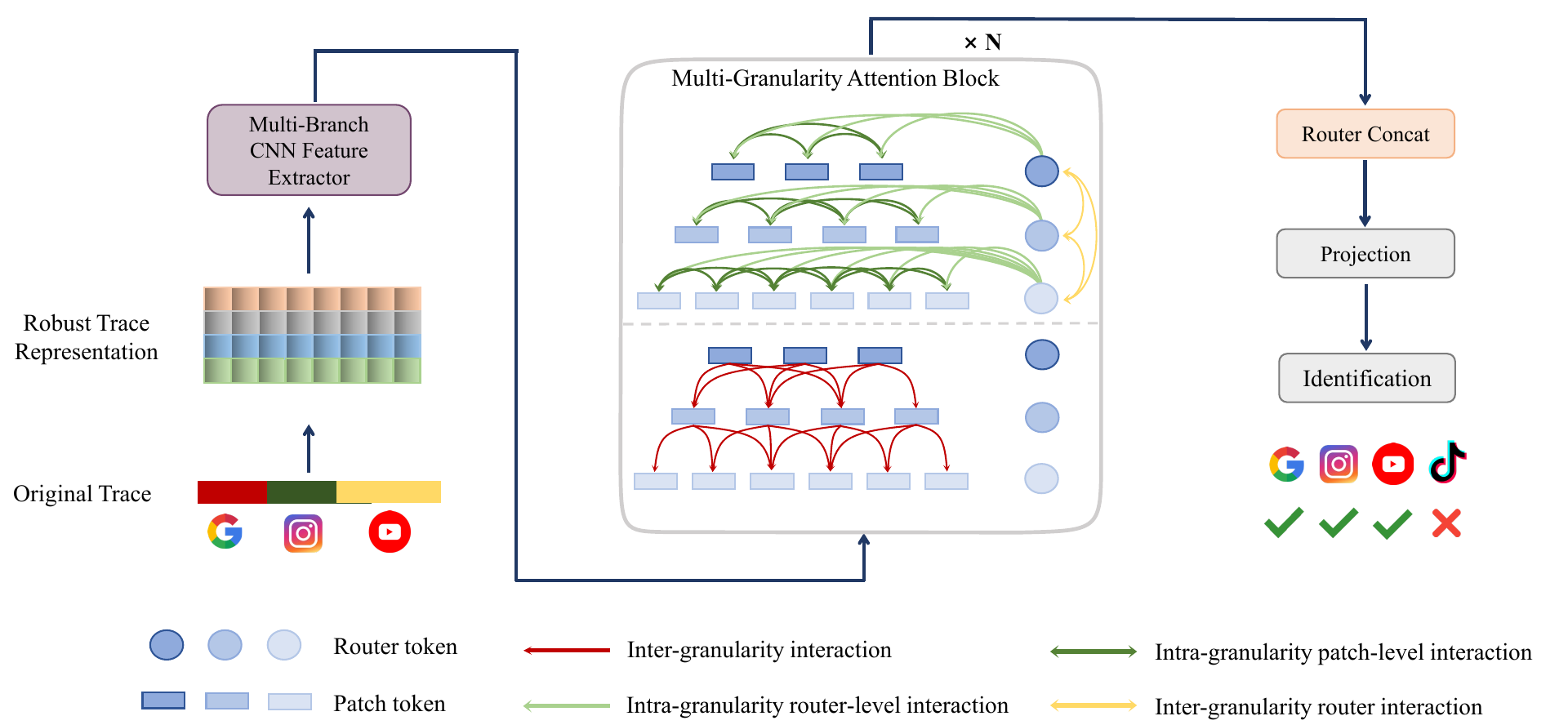}
    \caption{\added{Architecture of PrismWF. A raw mixed traffic trace is converted into a slot-based robust representation, encoded by multi-branch CNNs at multiple temporal granularities, refined through stacked Multi-Granularity Attention Blocks, and classified using the concatenated router tokens.}}
    \label{framework}
\end{figure*}

\section{Methodology}
\label{sec4}

In this section, we first introduce the overall architecture of PrismWF and then detail each component.
The pipeline of PrismWF is presented in Algorithm~\ref{alg:prismwf_pipeline}.

\subsection{Overall Architecture of PrismWF}
\added{As illustrated in Fig.~\ref{framework}, PrismWF first transforms raw website traffic into a robust trace representation by partitioning each trace into fixed-length time slots and extracting a six-channel feature matrix that captures both \emph{packet-level} statistics and \emph{temporal interval} characteristics.}
A multi-branch CNN feature extractor then models traffic patterns at different temporal granularities, where branches with distinct kernel sizes generate patch tokens covering coarse-grained contextual patterns and fine-grained local details.
To refine these multi-granularity representations, we introduce a traffic-aware Multi-Granularity Attention Block with a dedicated router token for each granularity.
Each block incorporates three complementary mechanisms: inter-granularity interaction for information exchange across temporal scales, intra-granularity interaction for patch-level semantic aggregation, and inter-granularity router interaction for fusing semantic summaries across granularities.
Finally, router tokens from all granularities are concatenated into a unified representation and fed into a linear classifier for multi-tab website identification.

\subsection{Robust Trace Representation}
Extracting effective traffic representations from interleaved website traces is critical for downstream WF attack models, as raw traffic is often contaminated by packets from concurrent websites and obfuscation mechanisms. In such settings, website-specific identity features are hard to isolate due to severe temporal noise and traffic interleaving. Existing representations based on global traffic statistics tend to overlook fine-grained temporal structures, making them vulnerable to WF defenses. 
Moreover, representations that rely solely on packet direction sequences fail to capture time-slot-level temporal dynamics, thereby limiting their ability to identify the injection timestamps of newly opened websites in multi-tab scenarios.
To address these challenges, we adopt a fixed-size time-slot strategy that abstracts raw packet events into structured local temporal segments. Specifically, the time axis of each traffic trace is partitioned into equal-length intervals with duration $\Delta t$, enabling localized modeling of traffic dynamics.

Let a raw traffic trace collected from the Tor network be denoted as
\begin{equation}
\mathbf{x} = \{ f_1, f_2, \dots, f_{N} \},
\end{equation}
where each packet event $f_i = \langle d_i, t_i \rangle$ consists of a packet direction $d_i$ and its corresponding timestamp $t_i$. 
Here, $d_i \in \{+1, -1\}$ indicates outgoing ($+1$) or incoming ($-1$) packets.
Accordingly, the traffic trace can be represented as a matrix $\mathbf{x} \in \mathbb{R}^{2 \times N}$, where the first row encodes packet directions and the second row records timestamp information. 
Given a predefined maximum page loading time $T$, the total number of time slots is computed as $L = \left\lceil \frac{T}{\Delta t} \right\rceil$.

For each time interval, we extract a six-dimensional feature vector to characterize the traffic pattern within that slot.
These features consist of two categories: packet-level statistics (4 dimensions) and time-interval features (2 dimensions).
Specifically, the packet-level statistics include (i) the numbers of incoming and outgoing packets, and (ii) the counts of direction transitions from outgoing to incoming and from incoming to outgoing packets.
The time-interval features capture the temporal gaps between consecutive outgoing-to-incoming and incoming-to-outgoing packet transitions within each time slot.
By aggregating these features across all time intervals, we construct a unified traffic feature matrix:
\begin{equation}
\mathbf{M}=\phi(\mathbf{x})\in\mathbb{R}^{6\times L},
\end{equation}
where $\phi(\cdot)$ denotes the traffic feature construction function. In practice, the actual number of intervals is truncated to $L$ if it exceeds the predefined value, or padded with zeros if it is insufficient to reach $L$. 
The detailed robust trace feature construction steps are presented in Algorithm \ref{alg:trace_feature}.

\begin{algorithm}[h]
\caption{Robust Trace Feature Construction $\mathbf{M}=\phi(\mathbf{x})$}
\label{alg:trace_feature}
\begin{algorithmic}[1]
  \Statex \textbf{Input:} Traffic trace $\mathbf{x}$; max loading time $T$; time slot size $\Delta t$
  \Statex \textbf{Output:} Feature matrix $\mathbf{M}\in\mathbb{R}^{6\times L}$

  \State $L \leftarrow \left\lceil \frac{T}{\Delta t} \right\rceil$ \Comment{Total time slots}
  \State $\mathbf{M}\leftarrow \mathbf{0}_{6\times L}$

  \For{$j = 1$ to $L$}
      \State $\mathcal{I}_j \leftarrow \{\, i \mid (j-1)\Delta t \le t_i < j\Delta t \,\}$ \Comment{Packet indices in $j$-th slot}
      \If{$\mathcal{I}_j=\emptyset$}
          \State \textbf{continue}
      \EndIf

      \State Extract ordered subsequence $\{(t_k,d_k)\}_{k=1}^{n}$ from $\{(t_i,d_i)\}_{i\in\mathcal{I}_j}$, where $n \leftarrow |\mathcal{I}_j|$
      \State $c^{+} \leftarrow \sum_{k=1}^{n}\mathbb{I}(d_k=+1)$,\quad $c^{-} \leftarrow \sum_{k=1}^{n}\mathbb{I}(d_k=-1)$
      \State $n^{+-}\leftarrow 0,\; n^{-+}\leftarrow 0,\; s^{+-}\leftarrow 0,\; s^{-+}\leftarrow 0$
      \For{$k = 1$ to $n-1$}
          \If{$d_k=+1 \land d_{k+1}=-1$}
              \State $n^{+-}\leftarrow n^{+-}+1$, $s^{+-}\leftarrow s^{+-}+(t_{k+1}-t_k)$
          \ElsIf{$d_k=-1 \land d_{k+1}=+1$}
              \State $n^{-+}\leftarrow n^{-+}+1$, $s^{-+}\leftarrow s^{-+}+(t_{k+1}-t_k)$
          \EndIf
      \EndFor
      \State $s^{+-}\leftarrow \mathbb{I}(n^{+-}>0)\cdot \frac{s^{+-}}{n^{+-}}$
      \State $s^{-+}\leftarrow \mathbb{I}(n^{-+}>0)\cdot \frac{s^{-+}}{n^{-+}}$

      \State $\mathbf{M}[:,j] \leftarrow [\,c^{+},c^{-},n^{+-},n^{-+},s^{+-},s^{-+}\,]^{\top}$
  \EndFor
  \State \Return $\mathbf{M}$
\end{algorithmic}
\end{algorithm}

\subsection{Multi-Granularity Feature Extraction}
Website traffic traces are long temporal sequences containing patterns at multiple time scales.
To capture traffic characteristics at different temporal resolutions, we employ $G$ parallel CNN branches with distinct kernel sizes to extract multi-granularity features from the robust traffic representation $\mathbf{M}$.
Each branch focuses on a specific temporal granularity and projects the extracted features into a shared embedding space.
For the $i$-th branch with kernel size $k_i$, a CNN feature extractor is applied to obtain:
\begin{equation}
\mathbf{F}_i=\mathrm{CNN}_i(\mathbf{M})\in\mathbb{R}^{d\times N_i},
\end{equation}
where $d$ denotes the unified embedding dimension and $N_i$ represents the number of patch tokens produced at the corresponding granularity.
\added{Each $\mathrm{CNN}_i$ comprises three stacked convolutional blocks following the DF feature-extractor design~\cite{sirinam2018deep}.}
Each ConvBlock includes two 1D convolutional layers with batch normalization and ReLU activation, followed by a max-pooling layer for temporal downsampling and a dropout layer for regularization.
Due to the use of different kernel sizes and pooling operations, each branch produces a feature sequence of length $N_i$, corresponding to a specific temporal granularity.

The final ConvBlock in each branch outputs $d$-dimensional features, thus ensuring all branches generate tokens in the pre-defined unified embedding space. We then transpose the feature matrix to obtain the patch token sequence for each granularity:
\begin{equation}
\mathbf{u}_i=\mathbf{F}_i^\top\in\mathbb{R}^{N_i\times d}.
\end{equation}
This design enables seamless interaction among tokens from different granularities in subsequent attention layers, while $N_i$ varies across branches due to the different downsampling rates induced by distinct kernel sizes.

\subsection{Multi-Granularity Attention Block}
After obtaining the multi-granularity representations of website traffic features, we further refine these features using the proposed Multi-Granularity Attention Block, which is designed to enable structured information interaction across temporal scales. 
The model stacks $B$ identical blocks, each consisting of three attention-based interaction layers that jointly support intra-granularity feature modeling and inter-granularity semantic aggregation.
Specifically, intra-granularity interactions capture temporal dependencies within each granularity, while inter-granularity interactions fuse complementary semantic information across different granularities. 
\added{Through this hierarchical attention-driven refinement, the model progressively improves the discriminability of traffic representations for website identification.}
The detailed design of each interaction layer is described below.

\noindent\textbf{Router Token Injection.}
To obtain compact and comparable representations across different temporal granularities, 
we introduce a learnable \emph{router token} for each granularity level. The router token serves as a granularity-level semantic proxy that summarizes patch-level features into a fixed-dimensional representation, which is amenable to downstream projection and classification.
\added{Specifically, for the $i$-th granularity, the initial learnable router token is defined as}
\begin{equation}
\added{\mathbf{r}_i^{(0)} \in \mathbb{R}^{1 \times d}.}
\end{equation}
\added{The router is carried separately during fine-to-coarse patch aggregation and concatenated with the corresponding patch sequence inside each intra-granularity hybrid attention operation. This design enables it to progressively aggregate granularity-specific semantic information without participating in cross-granularity patch interaction.}
\added{The resulting router tokens provide compact granularity-wise representations for subsequent cross-granularity router interaction and final classification.}

\noindent\textbf{Inter-granularity Interaction.}
\added{To exchange information across temporal scales, we perform fine-to-coarse aggregation, in which coarse-grained query tokens retrieve complementary information from finer-grained representations.}
\added{Traffic generated by concurrent page loads may dominate different portions of a trace or become temporally interleaved, producing mixed traffic patterns.}
Coarse-grained tokens provide a global and robust contextual perspective, while fine-grained tokens preserve detailed temporal variations. 
\added{The inter-granularity interaction mechanism uses coarse-grained context to query temporally aligned fine-grained features.}
During inter-granularity interaction, only patch tokens participate in cross-granularity information exchange, while router tokens are excluded.

Let $\mathbf{u}_i^{(b-1)}\in\mathbb{R}^{N_i\times d}$ denote the patch token sequence of the $i$-th granularity at the $(b-1)$-th Multi-Granularity Attention Block.
\added{For each pair of adjacent granularities, every coarse-grained patch token is mapped to a temporally corresponding local region in the finer-grained sequence, from which it retrieves complementary features through cross-attention.}
Assume the coarse-grained sequence contains $N_c$ patch tokens and the corresponding fine-grained sequence contains $N_f$ patch tokens.
For the $n$-th coarse-grained patch token (with zero-based indexing) acting as the query, we compute the center index of its corresponding region in the fine-grained sequence as
\begin{equation}
c_n = \left\lfloor (n + 0.5)\frac{N_f}{N_c} \right\rfloor,
\end{equation}
\added{where the offset $+0.5$ aligns the coarse-grained token with the temporal center rather than the left boundary of its corresponding fine-grained region.}

We then perform multi-head cross-attention (MHCA), which follows the standard attention formulation with queries and key--value pairs drawn from different token sequences. 
\added{Specifically, coarse-grained patch tokens act as queries, while fine-grained patch tokens serve as keys and values within a local temporal window that preserves alignment and limits contributions from distant positions.}
\added{Let $\mathbf{u}_c\in\mathbb{R}^{N_c\times d}$ denote the coarse-grained sequence and $\mathbf{u}_f\in\mathbb{R}^{N_f\times d}$ the fine-grained sequence.}
\added{For the $n$-th coarse-grained patch token, we select a local subset of fine-grained patch tokens centered at $c_n$. The odd-valued hyperparameter $w_{\mathrm{inter}}$ specifies the maximum number of fine-grained tokens attended by each coarse-grained token:}
\begin{equation}
\begin{aligned}
\mathbf{u}_f^{(c_n,w_{\mathrm{inter}})}
= \bigl\{
\mathbf{u}_f[k] \;\bigm|\;&
k \in \bigl[
\max\!\left(0,\, c_n-\left\lfloor \tfrac{w_{\mathrm{inter}}}{2} \right\rfloor\right), \\
&
\min\!\left(N_f-1,\, c_n+\left\lfloor \tfrac{w_{\mathrm{inter}}}{2} \right\rfloor\right)
\bigr]
\bigr\}.
\end{aligned}
\end{equation}

\added{Let $m_n$ denote the number of fine-grained tokens retained in $\mathbf{u}_f^{(c_n,w_{\mathrm{inter}})}$, such that $\mathbf{u}_f^{(c_n,w_{\mathrm{inter}})}\in\mathbb{R}^{m_n\times d}$. MHCA is then applied independently to each coarse-grained query token:}
\begin{equation}
\begin{aligned}
\added{\mathbf{u}_c'[n,:]}
&\added{=\mathrm{MHCA}\!\left(
\mathbf{u}_c[n,:],\,
\mathbf{u}_f^{(c_n,w_{\mathrm{inter}})},\,
\mathbf{u}_f^{(c_n,w_{\mathrm{inter}})}
\right)} \\
&\added{\in\mathbb{R}^{1\times d},\qquad n=0,\ldots,N_c-1.}
\end{aligned}
\end{equation}
\added{Applying this operation independently to all coarse-grained tokens and stacking the resulting vectors in temporal order gives $\mathbf{u}_c'=[\mathbf{u}_c'[0,:];\ldots;\mathbf{u}_c'[N_c-1,:]]\in\mathbb{R}^{N_c\times d}$.}
\added{Thus, each updated coarse-grained token incorporates temporally aligned fine-grained details while attending only to its corresponding local region.}

\noindent\textbf{Intra-granularity Interaction.}
\added{Within each temporal granularity, a hybrid self-attention operation jointly computes attention outputs for the patch tokens and router token. For block $b$, the patch tokens entering this stage are concatenated with the incoming router token:}
\begin{equation}
\added{\mathbf{X}_i^{(b)}=
[\mathbf{u}_i';\mathbf{r}_i^{(b-1)}]
\in\mathbb{R}^{(N_i+1)\times d}.}
\end{equation}
\added{The hybrid attention is computed as}
\begin{equation}
\added{\mathbf{H}_i^{(b)}=
\mathrm{MHA}_{\mathrm{hyb}}\!\left(
\mathbf{X}_i^{(b)},\mathbf{X}_i^{(b)},\mathbf{X}_i^{(b)};\mathbf{B}_i
\right),}
\end{equation}
\added{where $\mathbf{B}_i\in\mathbb{R}^{(N_i+1)\times(N_i+1)}$ is an additive bias applied to the scaled dot-product attention logits in every head. With zero-based query and key indices $p$ and $q$, respectively, and $\sigma=w_{\mathrm{intra}}/2$, its entries are}
\begin{equation}
\added{\mathbf{B}_i[p,q]=
\begin{cases}
-\dfrac{(p-q)^2}{2\sigma^2}, & p<N_i,\ q<N_i,\\
-\infty, & p<N_i,\ q=N_i,\\
0, & p=N_i,\ 0\leq q\leq N_i.
\end{cases}}
\end{equation}
\noindent\added{\textit{1) Patch-level local interaction:} For each patch query, the Gaussian distance bias favors temporally nearby patch tokens, while attention to the router token is masked. This interaction softly emphasizes local temporal dependencies among patch representations.}

\noindent\added{\textit{2) Router-level global aggregation:} For the router query, the attention bias is zero across all key positions. The router therefore attends globally to all patch tokens and itself, producing a granularity-level semantic summary.}

\added{The patch and router outputs are extracted from the same attention result as}
\begin{equation}
\added{\widehat{\mathbf{u}}_i^{(b)}=
\mathbf{H}_i^{(b)}[0{:}N_i,:],\qquad
\widehat{\mathbf{r}}_i^{(b)}=
\mathbf{H}_i^{(b)}[N_i,:].}
\end{equation}
\added{The intermediate router outputs $\widehat{\mathbf{r}}_i^{(b)}$ are passed to the subsequent inter-granularity router interaction, whereas the patch outputs $\widehat{\mathbf{u}}_i^{(b)}$ are used in the block-level residual update.}

\noindent\textbf{Inter-granularity Router Interaction.}
To enable global information exchange across different granularities within each Multi-Granularity Attention Block,
we perform global attention among router tokens from all granularities.
This design allows each router token to aggregate complementary global semantic information captured at other temporal scales.
\added{The intermediate routers from all granularities are concatenated into a router-level sequence}
\begin{equation}
\added{\widehat{\mathbf{R}}^{(b)} = [\widehat{\mathbf{r}}_1^{(b)}; \widehat{\mathbf{r}}_2^{(b)}; \dots; \widehat{\mathbf{r}}_G^{(b)}] \in \mathbb{R}^{G\times d},}
\end{equation}
\added{which is updated via global multi-head self-attention:}
\begin{equation}
\added{\mathbf{A}^{(b)} = 
\mathrm{MHA}_{\text{global}}\!\left(
\widehat{\mathbf{R}}^{(b)},\,
\widehat{\mathbf{R}}^{(b)},\,
\widehat{\mathbf{R}}^{(b)}
\right).}
\end{equation}

\added{The attention output $\mathbf{A}^{(b)}[i]$ is incorporated into the incoming router representation $\mathbf{r}_i^{(b-1)}$ through a residual connection, followed by a feed-forward update:}
\begin{align}
\added{\bar{\mathbf{r}}_i^{(b)}}
&\added{=\mathbf{r}_i^{(b-1)}+\mathrm{Dropout}\!\left(\mathbf{A}^{(b)}[i]\right),}
\label{eq:router_attention_residual}\\
\added{\mathbf{r}_i^{(b)}}
&\added{=\mathrm{LN}_{2}\!\left(\bar{\mathbf{r}}_i^{(b)}+
\mathrm{FFN}\!\left(\mathrm{LN}_{1}\!\left(\bar{\mathbf{r}}_i^{(b)}\right)\right)\right).}
\label{eq:router_ffn_residual}
\end{align}
\added{The patch outputs are combined with the incoming patch representations through an analogous residual feed-forward update. These residual paths preserve the incoming representations while incorporating features produced by the corresponding attention modules.}

\begin{algorithm}[!t]
\caption{\added{PrismWF Pipeline for Website Fingerprinting}}
\label{alg:prismwf_pipeline}
\begin{algorithmic}[1]
\Statex \textbf{Input:}  
Raw traffic trace $\mathbf{x}$; slot size $\Delta t$; maximum loading time $T$; 
branch kernels $\{k_i\}_{i=1}^{G}$; number of attention blocks $B$; 
\added{intra-granularity bias width $w_{\text{intra}}$; inter-granularity local window $w_{\text{inter}}$;} 
\added{(training only) multi-hot website label vector $\mathbf{y}$}
\Statex \textbf{Output:} \added{Predicted website confidence scores $\hat{\mathbf{y}}$}

\State \textbf{1) Robust Trace Representation}
\State $\mathbf{M} \leftarrow \phi(\mathbf{x}; \Delta t, T)$ 
\Comment{\added{$6\!\times\!L$ slot features}}

\State \textbf{2) Multi-Granularity Feature Extraction}
\For{$i = 1$ to $G$}
    \State $\mathbf{U}_i \leftarrow \mathrm{BranchCNN}_i(\mathbf{M}; k_i)$ 
    \Comment{\added{$N_i\!\times\!d$ patch tokens}}
    \State $\added{\mathbf{r}_i \equiv \mathbf{r}_i^{(0)}} \leftarrow \mathrm{InitRouter}(d)$ 
    \Comment{\added{$1\!\times\!d$ router token}}
\EndFor

\State \textbf{3) Stacked Multi-Granularity Attention Blocks}
\For{$b = 1$ to $B$}
{\color{black}
    \State $\mathcal{U}_{\mathrm{in}},\mathcal{R}_{\mathrm{in}}
    \leftarrow \{\mathbf{U}_i\}_{i=1}^{G},\{\mathbf{r}_i\}_{i=1}^{G}$
    \State \textbf{Fine-to-Coarse Patch Aggregation}
    \State $\mathcal{U}_{\mathrm{cross}} \leftarrow
    \mathrm{FineToCoarse}(\mathcal{U}_{\mathrm{in}},w_{\text{inter}})$

    \State \textbf{Intra-Granularity Hybrid Attention}
    \State $\widehat{\mathcal{U}},\widehat{\mathcal{R}} \leftarrow
    \mathrm{HybridIntra}(\mathcal{U}_{\mathrm{cross}},\mathcal{R}_{\mathrm{in}},w_{\text{intra}})$

    \State \textbf{Inter-Granularity Router Interaction}
    \State $\mathcal{A} \leftarrow \mathrm{RouterInteract}(\widehat{\mathcal{R}})$
    \For{$i = 1$ to $G$}
        \State $\mathbf{U}_i \leftarrow
        \mathrm{ResPostFFN}(\mathcal{U}_{\mathrm{in}}[i],\widehat{\mathcal{U}}[i])$
        \State $\mathbf{r}_i \leftarrow
        \mathrm{ResPostFFN}(\mathcal{R}_{\mathrm{in}}[i],\mathcal{A}[i])$
    \EndFor
}
\EndFor

\State \textbf{4) Website Identification}
\State $\mathbf{z} \leftarrow \mathrm{ConcatRouters}(\{\mathbf{r}_i\}_{i=1}^{G})$
\State $\added{\mathbf{o}} \leftarrow f(\mathbf{z})$
\Comment{\added{classification logits}}
\State $\added{\hat{\mathbf{y}} \leftarrow \mathrm{sigmoid}(\mathbf{o})}$
\Comment{\added{per-website confidence scores}}

\If{training}
    \State $\added{\mathcal{L} \leftarrow \mathrm{BCEWithLogits}(\mathbf{o},\mathbf{y})}$ 
    \State \textbf{update} model parameters by minimizing $\mathcal{L}$
\EndIf

\State \Return $\hat{\mathbf{y}}$
\end{algorithmic}
\end{algorithm}

\subsection{Website Identification}
After refining traffic representations across multiple temporal granularities via the router mechanism,
we perform website identification by aggregating global semantic information.
Specifically, the final router tokens from all granularities are concatenated to form a unified global traffic representation
$\mathbf{z} \in \mathbb{R}^{Gd}$, where each router token summarizes granularity-specific semantics produced by the last
Multi-Granularity Attention Block.
The global representation $\mathbf{z}$ is then fed into the website identification classifier $f$,
which applies a linear projection to produce classification logits $\mathbf{o}$ over $C$ website categories.
\added{Following the established formulation in prior multi-tab WF studies~\cite{guan2021bapm,jin2023transformer,deng2026towards}, PrismWF treats website identification from a mixed trace as a multi-label classification task. The corresponding confidence scores are $\hat{\mathbf{y}}=\mathrm{sigmoid}(\mathbf{o})$, and the model is optimized using binary cross-entropy with logits:}
\begin{equation}
\added{\mathcal{L}=\mathrm{BCEWithLogits}(\mathbf{o},\mathbf{y}),
\qquad \mathbf{y}\in\{0,1\}^{C},}
\end{equation}
\added{where $\mathbf{y}$ is the multi-hot ground-truth label vector.}

\section{Experiments}
In this section, we conduct extensive experiments on large-scale datasets to evaluate the proposed PrismWF.
We first describe the experimental setup in Section~\ref{sec51}, including datasets, evaluation metrics, baselines, and implementation details.
Section~\ref{sec52} evaluates PrismWF under both closed-world and open-world multi-tab WF scenarios.
\added{We further investigate the mixed-tab setting in Section~\ref{sec53}, assess statistical significance in Section~\ref{sec:significance}, and evaluate representative WF defenses in Section~\ref{sec54}.}
\added{We then evaluate PrismWF in a real-world Tor deployment in Section~\ref{sec:real_world} and examine computational efficiency in Section~\ref{sec56}.}
Finally, we conduct ablation studies in Section~\ref{sec55} to validate the effectiveness of each key component.


\label{sec5}

\subsection{Experimental Setup}
\label{sec51}

\subsubsection{Multi-tab Datasets}
We perform multi-tab WF attacks on the multi-tab datasets~\cite{deng2023robust,deng2026towards}, which model realistic multi-tab browsing behavior and include sub-datasets with different numbers of concurrently opened tabs (2-tab, 3-tab, 4-tab, and 5-tab).
In each sub-dataset, each traffic trace contains mixed traffic generated by the corresponding number of browser tabs.
ARES provides both closed-world and open-world evaluation scenarios.
In the closed-world setting, each sub-dataset contains traffic traces from 100 monitored website classes, with over 58{,}000 instances.
In the open-world setting, unmonitored traffic is additionally included while the number of monitored classes remains 100, resulting in 64{,}000 traffic instances per sub-dataset.
\added{The released ARES benchmark consists of directly collected multi-tab Tor sessions. Its collection tool initiates two to five homepage loads 3--10~s apart and records the aggregate traffic over a 240-s session. We follow the instance-level train, validation, and test split protocol implemented in WFlib.}\footnote{\url{https://github.com/FIND-Lab/Website-Fingerprinting-Library}}

\subsubsection{Multi-tab evaluation metrics}
For the multi-tab WF task, we follow prior work \cite{deng2023robust, deng2026towards, deng2025countmamba} and adopt Precision@K (P@K) and Mean Average Precision@K (MAP@K) to evaluate classification performance \cite{liu2017deep}. 
\added{We compute the ROC-AUC independently for each monitored website label and then average the resulting values across labels.}
Let $\mathbf{y}$ denote the ground-truth multi-tab vector associated with a traffic instance $x$, where $y_i = 1$ if $x$ contains traffic from the $i$-th website and $y_i = 0$ otherwise. 
Let $\hat{\mathbf{y}}$ denote the predicted confidence scores produced by the model over all monitored website labels.
Both P@K and MAP@K are computed based on the top-$K$ website labels ranked by prediction confidence. 
P@K measures the precision among the top-$K$ predicted website labels for each instance, defined as
\begin{equation}
\mathrm{P@K}(x) = \frac{1}{K} \sum_{i \in r_K(\hat{\mathbf{y}})} y_i,
\end{equation}
where $r_K(\hat{\mathbf{y}})$ denotes the set of website labels with the top-$K$ highest predicted confidence scores.
MAP@K extends P@K by further evaluating the ranking quality of predicted labels within the top-$K$ results. 
Specifically, it quantifies whether true website labels tend to rank higher than non-relevant ones by averaging the precision values at different cutoff positions:
\begin{equation}
\mathrm{MAP@K}(x) = \frac{1}{K} \sum_{i=1}^{K} \mathrm{P@i}(x).
\end{equation}
The final MAP@K score is obtained by averaging over all testing instances.

\setcounter{table}{1}
\begin{table*}[!t]
\centering
\color{black}
\caption{
\added{Comparison in closed-world and open-world multi-tab settings.}
}
\vspace{-0.2cm}
\label{tab:multi_tab_comparison}
\tiny
\setlength{\tabcolsep}{2.2pt}
\renewcommand{\arraystretch}{1.08}
\resizebox{\textwidth}{!}{
\begin{tabular}{l c c c c c c c c c c c c}
\toprule
Scenario & \# of tabs & Metrics & AWF & DF & Tik-Tok & Var-CNN & RF & BAPM & TMWF & CountMamba & ARES & PrismWF \\
\midrule
\multirow{8}{*}{Closed-world}
& \multirow{2}{*}{2-tab} & P@2 & 15.80$\pm$0.46 & 62.79$\pm$0.16 & 70.42$\pm$0.27 & 72.21$\pm$0.38 & 63.77$\pm$0.40 & 57.55$\pm$0.56 & 77.49$\pm$1.11 & 86.86$\pm$0.47 & 87.51$\pm$0.26 & \textbf{89.13$\pm$0.38} \\
&   & MAP@2 & 18.11$\pm$0.62 & 72.76$\pm$0.08 & 78.98$\pm$0.33 & 80.70$\pm$0.42 & 72.42$\pm$0.23 & 66.67$\pm$0.68 & 82.75$\pm$0.89 & 91.49$\pm$0.40 & 91.83$\pm$0.20 & \textbf{92.96$\pm$0.21} \\
\cmidrule(lr){2-13}
& \multirow{2}{*}{3-tab} & P@3 & 11.49$\pm$0.14 & 45.83$\pm$0.15 & 52.73$\pm$0.18 & 56.42$\pm$0.46 & 47.47$\pm$0.50 & 42.88$\pm$0.57 & 67.81$\pm$0.70 & 80.85$\pm$0.61 & 84.61$\pm$0.19 & \textbf{87.10$\pm$0.09} \\
&   & MAP@3 & 13.64$\pm$0.22 & 58.46$\pm$0.20 & 65.41$\pm$0.39 & 69.72$\pm$0.48 & 59.88$\pm$0.79 & 53.18$\pm$0.81 & 74.65$\pm$0.68 & 87.54$\pm$0.21 & 90.15$\pm$0.09 & \textbf{91.71$\pm$0.20} \\
\cmidrule(lr){2-13}
& \multirow{2}{*}{4-tab} & P@4 & 11.35$\pm$0.53 & 42.63$\pm$0.07 & 49.03$\pm$0.73 & 35.13$\pm$4.64 & 44.27$\pm$0.07 & 40.31$\pm$0.89 & 67.21$\pm$1.44 & 80.84$\pm$0.37 & 85.58$\pm$0.05 & \textbf{88.69$\pm$0.34} \\
&   & MAP@4 & 13.42$\pm$0.64 & 55.31$\pm$0.03 & 61.47$\pm$0.61 & 48.67$\pm$7.78 & 56.86$\pm$0.30 & 50.14$\pm$0.82 & 73.67$\pm$1.26 & 87.19$\pm$0.19 & 90.40$\pm$0.08 & \textbf{92.60$\pm$0.28} \\
\cmidrule(lr){2-13}
& \multirow{2}{*}{5-tab} & P@5 & 10.85$\pm$0.16 & 35.09$\pm$0.21 & 40.63$\pm$0.31 & 36.40$\pm$3.26 & 34.70$\pm$0.33 & 34.64$\pm$0.22 & 62.37$\pm$1.83 & 72.48$\pm$1.67 & 83.01$\pm$0.55 & \textbf{87.97$\pm$0.40} \\
&   & MAP@5 & 12.43$\pm$0.41 & 46.61$\pm$0.30 & 52.54$\pm$0.30 & 49.79$\pm$3.61 & 45.00$\pm$0.23 & 43.09$\pm$0.29 & 69.48$\pm$1.62 & 80.56$\pm$1.23 & 88.31$\pm$0.44 & \textbf{91.97$\pm$0.37} \\
\midrule
\multirow{8}{*}{Open-world}
& \multirow{2}{*}{2-tab} & P@2 & 17.76$\pm$0.29 & 60.53$\pm$0.21 & 68.27$\pm$0.29 & 70.61$\pm$0.54 & 62.72$\pm$0.13 & 55.77$\pm$0.95 & 74.08$\pm$0.53 & 85.04$\pm$0.02 & 86.01$\pm$0.42 & \textbf{87.90$\pm$0.12} \\
&   & MAP@2 & 20.58$\pm$0.29 & 70.41$\pm$0.23 & 77.00$\pm$0.36 & 79.37$\pm$0.32 & 71.59$\pm$0.16 & 64.93$\pm$0.92 & 80.11$\pm$0.44 & 90.12$\pm$0.07 & 90.75$\pm$0.25 & \textbf{91.97$\pm$0.09} \\
\cmidrule(lr){2-13}
& \multirow{2}{*}{3-tab} & P@3 & 12.26$\pm$0.18 & 45.01$\pm$0.15 & 52.51$\pm$0.34 & 57.40$\pm$0.87 & 47.80$\pm$0.35 & 41.89$\pm$0.74 & 68.58$\pm$1.82 & 80.38$\pm$0.69 & 83.96$\pm$0.29 & \textbf{86.72$\pm$0.61} \\
&   & MAP@3 & 14.50$\pm$0.14 & 58.31$\pm$0.09 & 65.58$\pm$0.32 & 70.92$\pm$0.73 & 60.79$\pm$0.27 & 52.47$\pm$0.99 & 75.69$\pm$1.83 & 87.47$\pm$0.34 & 89.74$\pm$0.25 & \textbf{91.69$\pm$0.29} \\
\cmidrule(lr){2-13}
& \multirow{2}{*}{4-tab} & P@4 & 12.08$\pm$0.43 & 42.04$\pm$0.23 & 47.76$\pm$0.54 & 41.38$\pm$3.62 & 43.99$\pm$0.56 & 39.60$\pm$0.62 & 66.57$\pm$0.71 & 79.10$\pm$0.98 & 84.89$\pm$0.22 & \textbf{88.22$\pm$0.21} \\
&   & MAP@4 & 14.46$\pm$0.57 & 54.60$\pm$0.26 & 60.45$\pm$0.56 & 55.23$\pm$4.40 & 56.54$\pm$0.48 & 49.36$\pm$0.92 & 73.20$\pm$0.67 & 85.81$\pm$0.61 & 89.96$\pm$0.06 & \textbf{92.22$\pm$0.13} \\
\cmidrule(lr){2-13}
& \multirow{2}{*}{5-tab} & P@5 & 11.76$\pm$0.42 & 36.08$\pm$0.42 & 41.88$\pm$0.09 & 36.18$\pm$3.56 & 36.92$\pm$0.57 & 35.32$\pm$0.64 & 63.97$\pm$2.24 & 74.13$\pm$1.29 & 83.84$\pm$0.46 & \textbf{88.58$\pm$0.31} \\
&   & MAP@5 & 13.70$\pm$0.47 & 48.13$\pm$0.39 & 54.28$\pm$0.25 & 48.86$\pm$5.29 & 47.69$\pm$0.71 & 44.34$\pm$0.63 & 71.10$\pm$1.72 & 81.88$\pm$1.07 & 89.03$\pm$0.25 & \textbf{92.33$\pm$0.15} \\
\bottomrule
\end{tabular}}
\end{table*}

\setcounter{table}{0}
\begin{table}[!t]
\centering
\caption{\added{Default configuration of PrismWF.}}
\label{tab:hg_params}

\footnotesize
\renewcommand{\arraystretch}{1.12}

\begin{tabular*}{\columnwidth}{
@{\extracolsep{\fill}}
>{\centering\arraybackslash}m{2.25cm}
>{\centering\arraybackslash}m{3.75cm}
>{\centering\arraybackslash}m{1.65cm}
@{}
}
\toprule
\textbf{Model Part} & \textbf{Details} & \textbf{Value} \\
\midrule

\multirow{5}{*}{Multi-Branch CNN}
& Embedding Dimension & 256 \\
& Branch Number & 4 \\
& Kernel Sizes & [15, 11, 7, 5] \\
& ConvBlocks per Branch & 3 \\
& \added{Patch Tokens ($L=8{,}000$)}
& \added{[3, 7, 24, 64]} \\

\midrule

\multirow{7}{*}{\shortstack{Multi-Granularity\\Block}}
& Block Number & 3 \\
& Attention Head Number & 8 \\
& \added{Intra-Granularity Bias Width} & 5 \\
& Inter-Granularity Window & 3 \\
& FFN Dimension & 1024 \\
& \added{Dropout Rate} & \added{0.1} \\
& \added{Residual / Normalization}
& \added{Residual + LayerNorm} \\

\midrule

Classification Head
& Router Fusion Dimension
& $256 \times 4$ \\

\midrule

\added{Input Processing}
& \added{Feature Normalization}
& \added{None} \\

\bottomrule
\end{tabular*}
\end{table}
\setcounter{table}{2}

\begin{table*}[!t]
\centering
\color{black}
\caption{\added{Oracle-$K$ performance under mixed-tab training and fixed-tab testing.}}
\vspace{-0.2cm}
\label{tab:multitab_test_tabs}
\scriptsize
\setlength{\tabcolsep}{3.5pt}
\renewcommand{\arraystretch}{1.15}

\resizebox{0.90\textwidth}{!}{%
\begin{tabular}{l | cc | cc | cc | cc}
\toprule
\textbf{Method}
& \multicolumn{2}{c|}{\textbf{2-tab (Test)}}
& \multicolumn{2}{c|}{\textbf{3-tab (Test)}}
& \multicolumn{2}{c|}{\textbf{4-tab (Test)}}
& \multicolumn{2}{c}{\textbf{5-tab (Test)}} \\
\cline{2-9}
 & P@2 & MAP@2 & P@3 & MAP@3 & P@4 & MAP@4 & P@5 & MAP@5 \\
\midrule
TMWF
& 59.05$\pm$1.26 & 69.40$\pm$0.61
& 46.86$\pm$1.28 & 60.23$\pm$1.31
& 43.63$\pm$1.22 & 56.69$\pm$1.14
& 37.65$\pm$0.79 & 49.65$\pm$1.14 \\

CountMamba
& 78.02$\pm$0.20 & 86.28$\pm$0.18
& 67.77$\pm$0.29 & 80.77$\pm$0.36
& 63.92$\pm$0.60 & 78.03$\pm$0.33
& 54.03$\pm$1.17 & 69.35$\pm$0.73 \\

ARES
& 80.43$\pm$0.46 & 87.71$\pm$0.34
& 72.68$\pm$0.01 & 83.89$\pm$0.07
& 70.70$\pm$0.77 & 82.47$\pm$0.40
& 63.40$\pm$0.85 & 76.60$\pm$0.59 \\

PrismWF
& \textbf{82.35$\pm$0.36} & \textbf{89.05$\pm$0.26}
& \textbf{75.75$\pm$0.23} & \textbf{85.97$\pm$0.23}
& \textbf{74.70$\pm$0.43} & \textbf{85.29$\pm$0.36}
& \textbf{67.81$\pm$0.36} & \textbf{79.83$\pm$0.19} \\
\bottomrule
\end{tabular}
}

\end{table*}

\begin{table}[!t]
\centering
\color{black}
\caption{\added{Unknown-$K$ performance after mixed-tab training.}}
\label{tab:unknown_k}
\scriptsize
\setlength{\tabcolsep}{3pt}
\renewcommand{\arraystretch}{1.08}
\resizebox{\columnwidth}{!}{
\begin{tabular}{l c c c c c}
\toprule
\textbf{Method} & \textbf{Micro-Prec. (\%)} & \textbf{Micro-Rec. (\%)} & \textbf{Micro-F1 (\%)} & \textbf{Exact (\%)} & \textbf{Count MAE $\downarrow$} \\
\midrule
TMWF & 55.59$\pm$1.51 & 38.87$\pm$0.95 & 45.75$\pm$1.05 & 7.75$\pm$0.86 & 1.39$\pm$0.05 \\
CountMamba & 81.76$\pm$0.44 & 56.03$\pm$0.64 & 66.50$\pm$0.50 & 23.23$\pm$0.53 & 1.23$\pm$0.01 \\
ARES & 85.16$\pm$1.27 & 62.43$\pm$1.43 & 72.03$\pm$0.56 & 30.25$\pm$1.27 & 1.05$\pm$0.07 \\
PrismWF & \textbf{85.60$\pm$0.70} & \textbf{66.96$\pm$0.08} & \textbf{75.14$\pm$0.30} & \textbf{35.01$\pm$0.40} & \textbf{0.90$\pm$0.01} \\
\bottomrule
\end{tabular}}
\end{table}

\subsubsection{Baselines}
\added{We compare PrismWF with five representative single-tab WF attacks, namely AWF~\cite{rimmer2017automated}, DF~\cite{sirinam2018deep}, Tik-Tok~\cite{rahman2019tik}, Var-CNN~\cite{bhat2018var}, and RF~\cite{shen2023subverting}, as well as four multi-tab WF attacks: BAPM~\cite{guan2021bapm}, TMWF~\cite{jin2023transformer}, CountMamba~\cite{deng2025countmamba}, and ARES~\cite{deng2023robust,deng2026towards}.}
\added{For traditional single-tab attacks, we remove the softmax layer, retain one logit per website label, and train the adapted classifier using the multi-label soft-margin loss \cite{deng2023robust, deng2025countmamba, deng2026towards}.}
\added{All attacks use the same train, validation, and test partitions. The baseline methods retain their original or released preprocessing pipelines and input representations, while PrismWF uses the six-channel robust trace representation described in Section~\ref{sec4}. The adapted single-tab baselines retain the method-specific feature extractors, backbone architectures, and optimization settings used in WFlib, including the optimizer, learning rate, batch size, and learning-rate schedule. All configurations are fixed before test evaluation without tuning on any test set. Fixed-tab checkpoints maximize validation MAP@K, whereas mixed-tab checkpoints maximize pooled validation MAP@5.}
\label{single-tab setting}

\subsubsection{Implementation detail}
\added{We train PrismWF using Python~3.10.19 and PyTorch~2.1.2.}
\added{All experiments are conducted on NVIDIA A800-SXM4-80GB GPUs, and all attack models are trained for 80 epochs.}
\added{Unless otherwise stated, all repeated experiments use random seeds 2024, 2025, and 2026, and results are reported as the mean $\pm$ sample standard deviation.}
More detailed deployment configurations of PrismWF are provided in Table~\ref{tab:hg_params}.
\added{PrismWF uses AdamW with an initial learning rate of $5\times10^{-4}$, a batch size of 256, and StepLR with a step size of 30 and a decay factor of 0.74. Traces are right-truncated or zero-padded to 8{,}000 slots. Unknown-$K$ thresholds are selected only on the pooled validation sets as described in Section~\ref{sec53}.}
\added{The public repository also provides data-preparation and evaluation scripts, reproducible training configurations, and the custom WFDefProxy transport used for online DeTorrent deployment. For datasets subject to redistribution restrictions, it provides deterministic preparation scripts and official download instructions instead of repackaging protected data.}

\subsection{Multi-Tab Attack Performance}
\label{sec52}

In this section, we evaluate the performance of WF attack models in multi-tab scenarios under both closed-world and open-world settings.
For baseline methods originally designed for single-tab WF (e.g., AWF, DF, Tik-Tok, Var-CNN, and RF), we adopt the same multi-tab adaptation strategy described in Section~\ref{single-tab setting}, following the standard practice of prior multi-tab WF attacks.
For representative multi-tab WF methods (BAPM, TMWF, ARES, and CountMamba), we directly use their publicly released codes for training and testing without modifications.

\subsubsection{Closed-world scenario}
\added{As shown in Table~\ref{tab:multi_tab_comparison}, PrismWF consistently achieves the best P@K and MAP@K across all closed-world settings. In the 2-tab setting, PrismWF obtains 89.13\% P@2 and 92.96\% MAP@2, exceeding ARES by 1.62 and 1.13 percentage points, respectively. Under the more challenging 5-tab setting, PrismWF achieves 87.97\% P@5 and 91.97\% MAP@5, outperforming ARES by 4.96 and 3.66 percentage points and CountMamba by 15.49 and 11.41 percentage points, respectively. Notably, the performance advantage becomes more pronounced as the number of concurrent tabs increases, indicating that PrismWF is better able to handle the stronger traffic interleaving and feature mixing introduced by multi-tab browsing. The low standard deviations further indicate stable performance across repeated training runs.}

\subsubsection{Open-world scenario}
Consistent with prior work \cite{sirinam2018deep, jin2023transformer, deng2025countmamba, deng2023robust}, all unmonitored websites are grouped into a single class—introducing an additional label compared to the closed-world setting and substantially increasing classification difficulty, especially under multi-tab traffic mixing.
This setting reflects a more realistic, challenging threat model, as attackers must handle unknown background traffic.

\added{PrismWF also consistently outperforms all baselines in the open-world setting. In the 2-tab setting, it achieves 87.90\% P@2 and 91.97\% MAP@2. Under the more challenging 5-tab setting, PrismWF obtains 88.58\% P@5 and 92.33\% MAP@5, exceeding ARES by 4.74 and 3.30 percentage points, respectively. Its advantage over CountMamba further increases to 14.45 percentage points in P@5 and 10.45 percentage points in MAP@5. These results indicate that the proposed multi-granularity representation learning and router-based aggregation remain effective under the combined effects of unmonitored traffic and severe inter-tab mixing.}

\begin{figure}[!t]
    \centering
    \includegraphics[width=0.96\columnwidth]{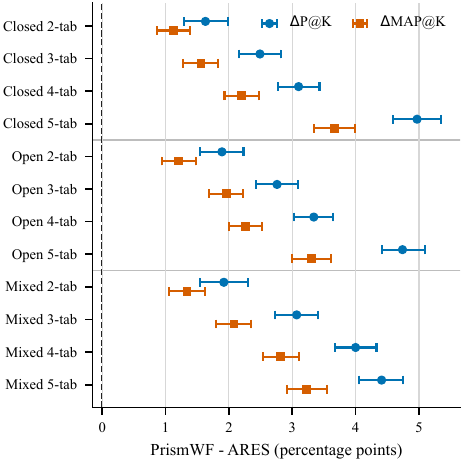}
    \caption{\added{Paired bootstrap comparison between PrismWF and ARES. Markers show the mean PrismWF-minus-ARES difference, and horizontal bars denote percentile-based 95\% confidence intervals. All differences are reported in percentage points, and positive values favor PrismWF.}}
    \label{fig:significance}
\end{figure}

\subsection{\texorpdfstring{\textcolor{black}{Mixed-Tab Training and Unknown Tab Counts}}{Mixed-Tab Training and Unknown Tab Counts}}
\label{sec53}


\added{To evaluate model performance more realistically, we consider an attacker without prior knowledge of the tab count. Separate fixed-tab models would therefore require prior tab-count estimation and model selection. We instead sample 30\% of the traffic traces from each of the 2-tab, 3-tab, 4-tab, and 5-tab datasets and merge them into a common set for \textit{mixed-tab training}. This protocol exposes one model to varying traffic-interleaving patterns across 2--5-tab sessions. For comparability with prior work, we first report conventional oracle-$K$ ranking performance on each fixed-tab test set, where the true tab count is used only to define the evaluation cutoff. We then remove this evaluation-time assumption in the unknown-$K$ evaluation below.}

\added{As shown in Table~\ref{tab:multitab_test_tabs}, PrismWF consistently achieves the best performance across all testing tab settings. On the 2-tab test set, PrismWF attains 82.35\% P@2 and 89.05\% MAP@2, outperforming ARES by 1.92 and 1.34 percentage points, respectively.}
More importantly, PrismWF maintains strong performance advantages under heavier traffic mixing.
\added{On the most challenging 5-tab test set, PrismWF achieves 67.81\% P@5 and 79.83\% MAP@5, exceeding CountMamba by 13.78 and 10.48 percentage points and ARES by 4.41 and 3.23 points, respectively.}
We observe a consistent decline in P@K and MAP@K as the number of concurrent tabs increases, as expected: identifying all target websites grows more difficult with stronger temporal interleaving and feature mixing in traffic traces.
\added{Nevertheless, PrismWF maintains a clear advantage over the evaluated baselines, suggesting that its multi-granularity representation learning and router-based interaction mechanisms remain effective under severe traffic mixing and heterogeneous tab counts.}

\added{Beyond conventional oracle-$K$ ranking, we evaluate a practical unknown-$K$ setting in which the attacker does not know the tab count. We select one global probability threshold for each trained model by maximizing micro-F1 on the pooled 2--5-tab validation sets. At test time, the model predicts every website whose sigmoid probability exceeds this threshold. No true $K$ value or top-$K$ fallback is used. We pool the four test sets, comprising 23{,}200 traces, and report five complementary metrics. Micro-precision measures the fraction of predicted website labels that are correct, while micro-recall measures the fraction of ground-truth website labels that are recovered. Micro-F1 is their harmonic mean. Exact-set accuracy counts a trace as correct only when its predicted and ground-truth label sets are identical. Count MAE averages the absolute difference between the predicted and true numbers of labels, for which a lower value is better.}

\added{Table~\ref{tab:unknown_k} shows that PrismWF remains the best-performing method without access to the true tab count. It improves micro-F1 over ARES by 3.11 points and exact-set accuracy by 4.76 points, while reducing count MAE from 1.05 to 0.90. Lower recall than precision across all methods indicates that unknown-$K$ prediction remains challenging.}

\subsection{\texorpdfstring{\textcolor{black}{Statistical Significance}}{Statistical Significance}}
\label{sec:significance}

\added{To further quantify the uncertainty in the observed improvements, we conduct a paired bootstrap comparison between PrismWF and ARES. We select ARES because it is the strongest competing method in every closed-world, open-world, and mixed-tab setting reported in Tables~\ref{tab:multi_tab_comparison} and~\ref{tab:multitab_test_tabs}, making it the most conservative baseline for this analysis. For each method, we first average the prediction scores of every test trace across seeds 2024, 2025, and 2026. We then compute trace-level P@K and MAP@K from these seed-averaged predictions and form the paired PrismWF-minus-ARES differences. Finally, we draw 10{,}000 bootstrap resamples from the paired trace-level differences and report percentile-based 95\% confidence intervals. This analysis complements the across-seed standard deviations by quantifying uncertainty arising from finite test-set sampling.}

\added{As shown in Fig.~\ref{fig:significance}, all percentile-based 95\% confidence intervals exclude zero, indicating consistently positive PrismWF improvements under test-set resampling. The P@K advantage increases from 1.63 to 4.97 percentage points as the tab count grows in the closed world. It similarly increases from 1.89 to 4.74 points in the open world and from 1.92 to 4.41 points under mixed-tab training. The three-seed means and standard deviations characterize variation across training runs, while the paired intervals quantify uncertainty under test-set resampling. 
Together, these results provide consistent evidence that PrismWF improves upon the strongest baseline across different tab counts and evaluation settings.}

\subsection{Multi-Tab Attack against WF Defenses}
\label{sec54}

In this section, we select three representative WF defense mechanisms (WTF-PAD~\cite{juarez2016toward}, Front~\cite{gong2020zero}, and RegulaTor~\cite{holland2020regulator}) to evaluate their effectiveness against WF attacks in multi-tab browsing scenarios.
Tamaraw~\cite{cai2014systematic} is excluded due to its prohibitively high bandwidth overhead and latency for practical deployment.
WTF-PAD adaptively injects dummy packets into raw traffic traces to perturb original website traffic, disrupting salient statistical features exploited by fingerprinting attacks.
Front targets information leakage in the initial phase of website access and adopts a Rayleigh-distribution-based random dummy packet padding strategy to obfuscate front-end traffic patterns.
RegulaTor combines packet delay manipulation with dummy packet injection to effectively blur burst-level statistical characteristics of website traffic.
We evaluate multi-tab WF attacks under these defenses in two representative settings: a 2-tab scenario and a more challenging 5-tab scenario.
As shown in Table~\ref{tab:defense_2tab_pmap} and~\ref{tab:defense_5tab_pmap}, PrismWF consistently achieves state-of-the-art performance across all six dataset–defense combinations.

\added{Table~\ref{tab:defense_2tab_pmap} reports the 2-tab defense results. PrismWF achieves the highest P@2 and MAP@2 under WTF-PAD, Front, and RegulaTor. Its P@2 values are 84.95\%, 87.01\%, and 71.55\%, respectively, while the corresponding MAP@2 values are 89.77\%, 91.46\%, and 77.67\%. CountMamba falls to approximately 2.8\% under the simulated RegulaTor traces, showing severe degradation in this evaluation setting.}

\begin{table}[H]
\centering
\color{black}
\caption{\added{Performance against simulated defenses in the closed-world 2-tab setting.}}
\label{tab:defense_2tab_pmap}

\scriptsize
\setlength{\tabcolsep}{3.2pt}
\renewcommand{\arraystretch}{1.08}

\resizebox{\columnwidth}{!}{
\begin{tabular}{l | cc | cc | cc}
\toprule
 & \multicolumn{2}{c|}{\textbf{WTF-PAD}} 
 & \multicolumn{2}{c|}{\textbf{Front}} 
 & \multicolumn{2}{c}{\textbf{RegulaTor}} \\
\cline{2-7}
\textbf{Attack} 
& \textbf{P@2} & \textbf{MAP@2}
& \textbf{P@2} & \textbf{MAP@2}
& \textbf{P@2} & \textbf{MAP@2} \\
\midrule
TMWF        & 61.36$\pm$1.18 & 68.21$\pm$1.28 & 64.55$\pm$1.56 & 72.66$\pm$1.46 & 40.77$\pm$0.66 & 47.24$\pm$0.85 \\
CountMamba  & 78.84$\pm$0.60 & 85.17$\pm$0.38 & 83.81$\pm$0.14 & 89.23$\pm$0.10 & 2.78$\pm$0.08 & 2.76$\pm$0.07 \\
ARES        & 82.60$\pm$0.17 & 87.91$\pm$0.12 & 85.16$\pm$0.34 & 90.08$\pm$0.27 & 70.55$\pm$0.47 & 77.02$\pm$0.48 \\
PrismWF     & \textbf{84.95$\pm$0.32} & \textbf{89.77$\pm$0.16} 
            & \textbf{87.01$\pm$0.25} & \textbf{91.46$\pm$0.24}
            & \textbf{71.55$\pm$0.57} & \textbf{77.67$\pm$0.56} \\
\bottomrule
\end{tabular}
}
\end{table}

\added{Table~\ref{tab:defense_5tab_pmap} shows that stronger traffic interleaving reduces the performance of every attack under the three defenses. PrismWF remains the best-performing method, achieving 79.93\%, 83.31\%, and 53.26\% P@5 under WTF-PAD, Front, and RegulaTor, respectively. The corresponding MAP@5 values are 85.55\%, 88.46\%, and 59.55\%. Compared with ARES, PrismWF improves P@5 by 5.66, 5.14, and 0.38 percentage points under the three defenses, respectively. These results demonstrate that PrismWF preserves its advantage under both stronger traffic mixing and defense-induced perturbations.}

\begin{table}[H]
\centering
\color{black}
\caption{\added{Performance against simulated defenses in the closed-world 5-tab setting.}}
\label{tab:defense_5tab_pmap}

\scriptsize
\setlength{\tabcolsep}{3.2pt}
\renewcommand{\arraystretch}{1.08}

\resizebox{\columnwidth}{!}{
\begin{tabular}{l | cc | cc | cc}
\toprule
 & \multicolumn{2}{c|}{\textbf{WTF-PAD}} 
 & \multicolumn{2}{c|}{\textbf{Front}} 
 & \multicolumn{2}{c}{\textbf{RegulaTor}} \\
\cline{2-7}
\textbf{Attack} 
& \textbf{P@5} & \textbf{MAP@5}
& \textbf{P@5} & \textbf{MAP@5}
& \textbf{P@5} & \textbf{MAP@5} \\
\midrule
TMWF        & 44.18$\pm$1.50 & 51.48$\pm$1.59 & 41.06$\pm$2.86 & 48.84$\pm$2.70 & 17.50$\pm$1.71 & 20.16$\pm$1.87 \\
CountMamba  & 59.18$\pm$0.70 & 68.39$\pm$0.65 & 65.33$\pm$1.47 & 74.40$\pm$1.05 & 5.69$\pm$0.04 & 5.78$\pm$0.08 \\
ARES        & 74.27$\pm$0.35 & 80.90$\pm$0.25 & 78.17$\pm$0.46 & 84.48$\pm$0.47 & 52.88$\pm$0.70 & 59.33$\pm$0.67 \\
PrismWF     & \textbf{79.93$\pm$1.84} & \textbf{85.55$\pm$1.43} 
            & \textbf{83.31$\pm$0.64} & \textbf{88.46$\pm$0.44}
            & \textbf{53.26$\pm$0.22} & \textbf{59.55$\pm$0.12} \\
\bottomrule
\end{tabular}
}
\end{table}

\subsection{\texorpdfstring{\textcolor{black}{Real-World Deployment Evaluation}}{Real-World Deployment Evaluation}}
\label{sec:real_world}

\added{To evaluate PrismWF under real-world deployment conditions, we built a live Tor testbed using Tor Browser and WFDefProxy~\cite{gong2023wfdefproxy} and collected four datasets in August 2026. The datasets cover 2-tab and 5-tab sessions under undefended and online DeTorrent conditions. The collection includes 96 contemporary websites with Tranco ranks ranging from 6 to 500, spanning social platforms, streaming services, e-commerce sites, news sites, and web applications. Two-tab sessions initiate the two page loads 3--10 s apart, while 5-tab sessions independently sample a 3--10 s interval between consecutive page loads. Each session uses a 160-s observation window followed by a 5-s period to drain in-flight traffic. WFDefProxy records the transmitted traffic in fixed 536-byte frames, and only cells within the observation window are converted into model inputs. The undefended condition uses the WFDefProxy null transport. For the defended condition, we reproduced a DeTorrent generator checkpoint using the official training code and dataset without modifying the defense model, then integrated the generator into a custom WFDefProxy transport. Both tab settings use the same generator checkpoint without retraining.}

\added{Each dataset is independently split using an 8:1:1 ratio, and all attacks use identical partitions within that dataset. Because the undefended and defended traces are collected in separate live Tor sessions, they may experience different website states and network conditions. We therefore compare attacks within each condition and do not attribute cross-condition differences solely to DeTorrent.}

\begin{table}[!t]
\centering
\color{black}
\caption{\added{Performance in the real-world 2-tab and 5-tab Tor deployments.}}
\label{tab:real_world_deployment}
\scriptsize
\setlength{\tabcolsep}{3.6pt}
\renewcommand{\arraystretch}{1.08}
\resizebox{\columnwidth}{!}{
\begin{tabular}{c l l c c}
\toprule
\textbf{Tabs} & \textbf{Condition} & \textbf{Method} & \textbf{P@$K$} & \textbf{MAP@$K$} \\
\midrule
\multirow{4}{*}{2}
& \multirow{2}{*}{Undefended} & ARES    & $88.79{\pm}0.45$ & $93.45{\pm}0.35$ \\
&                              & PrismWF & $\mathbf{89.79{\pm}0.46}$ & $\mathbf{94.20{\pm}0.34}$ \\
\cmidrule(lr){2-5}
& \multirow{2}{*}{Online DeTorrent} & ARES    & $84.08{\pm}0.25$ & $90.66{\pm}0.37$ \\
&                                    & PrismWF & $\mathbf{85.01{\pm}0.33}$ & $\mathbf{91.40{\pm}0.08}$ \\
\midrule
\multirow{4}{*}{5}
& \multirow{2}{*}{Undefended} & ARES    & $64.99{\pm}0.57$ & $\mathbf{78.97{\pm}0.47}$ \\
&                              & PrismWF & $\mathbf{65.20{\pm}0.66}$ & $78.41{\pm}0.83$ \\
\cmidrule(lr){2-5}
& \multirow{2}{*}{Online DeTorrent} & ARES    & $49.17{\pm}0.40$ & $59.78{\pm}0.21$ \\
&                                    & PrismWF & $\mathbf{54.77{\pm}0.95}$ & $\mathbf{64.73{\pm}0.90}$ \\
\bottomrule
\end{tabular}}
\end{table}

\added{Table~\ref{tab:real_world_deployment} reports the results for each real-world collection condition. PrismWF achieves the highest P@2 and MAP@2 in both 2-tab conditions. Without defense, its gains over ARES are 1.00 percentage points in P@2 and 0.75 points in MAP@2; under online DeTorrent, the corresponding gains are 0.93 and 0.74 points. In the undefended 5-tab setting, PrismWF improves P@5 by 0.21 points, while ARES leads MAP@5 by 0.56 points, indicating comparable overall ranking performance. Under online DeTorrent in the 5-tab setting, PrismWF improves P@5 and MAP@5 by 5.60 and 4.95 points, respectively. These within-dataset comparisons show that PrismWF remains effective on recently collected multi-tab Tor traffic under both undefended and online defended conditions.}

\subsection{\texorpdfstring{\textcolor{black}{Computational Efficiency}}{Computational Efficiency}}
\label{sec56}

\added{We compare the computational cost of PrismWF with three representative multi-tab WF attacks under the closed-world 2-tab and 5-tab settings. Table~\ref{tab:efficiency} summarizes model complexity, training cost, and inference efficiency. All models are trained and deployed for inference on an NVIDIA A800-SXM4-80GB GPU. FLOPs are profiled per trace, and training time is measured over 80 epochs. Single-trace latency and peak GPU memory are measured with a batch size of one, whereas throughput is measured with a batch size of 64.}

\begin{table*}[!t]
\centering
\color{black}
\caption{\added{Computational efficiency in the closed-world 2-tab and 5-tab settings.}}
\label{tab:efficiency}
\scriptsize
\setlength{\tabcolsep}{4.5pt}
\renewcommand{\arraystretch}{1.08}
\resizebox{\textwidth}{!}{
\begin{tabular}{c l c c c c c c c}
\toprule
\textbf{Tabs} & \textbf{Method} & \textbf{Params. (M)} & \textbf{Size (MB)} & \textbf{FLOPs (G)} & \textbf{Train (min)} & \textbf{Latency (ms)} & \textbf{Throughput (trace/s)} & \textbf{Memory (MB)} \\
\midrule
\multirow{4}{*}{2}
& ARES       & 4.15  & 15.88 & 0.187 & \textbf{27.87} & \textbf{3.03$\pm$0.05} & \textbf{11700.25} & 153.70 \\
& TMWF       & 4.85  & 18.56 & 0.498 & 62.94 & 3.10$\pm$0.08 & 4876.86 & \textbf{42.43} \\
& CountMamba & \textbf{2.45} & \textbf{9.39} & 0.088 & 250.59 & 4.63$\pm$0.33 & 5595.82 & 161.65 \\
& PrismWF    & 22.69 & 86.73 & 0.295 & 116.65 & 28.12$\pm$0.85 & 1623.37 & 231.42 \\
\midrule
\multirow{4}{*}{5}
& ARES       & 4.15  & 15.88 & 0.187 & \textbf{29.28} & \textbf{3.02$\pm$0.07} & \textbf{11481.92} & 153.70 \\
& TMWF       & 4.85  & 18.57 & 0.510 & 62.28 & 3.07$\pm$0.07 & 4703.70 & \textbf{42.43} \\
& CountMamba & \textbf{2.45} & \textbf{9.39} & 0.088 & 351.44 & 4.52$\pm$0.14 & 5393.58 & 161.65 \\
& PrismWF    & 22.69 & 86.73 & 0.295 & 117.84 & 27.84$\pm$0.51 & 1610.93 & 231.42 \\
\bottomrule
\end{tabular}}
\end{table*}

\added{As shown in Table~\ref{tab:efficiency}, PrismWF incurs a relatively high computational cost, with 22.69 million parameters and 0.295 GFLOPs per trace. However, its computational cost remains essentially unchanged as the number of tabs increases: the single-trace latency is 28.12 ms and 27.84 ms for the 2-tab and 5-tab settings, respectively, while the corresponding throughputs are 1623.37 and 1610.93 traces/s. This stability results from the fixed-length trace representation and the absence of tab-specific query tokens. Training time is also similar across the two settings, at 116.65 and 117.84 minutes. Overall, PrismWF requires additional computation for improved recognition performance, but its cost remains stable as the tab count increases.}

\subsection{Ablation Study}
\label{sec55}

In this section, we conduct systematic ablation studies on the proposed model to evaluate the contributions of each core component to the overall WF attack performance, and further investigate the sensitivity of key hyperparameters.

\subsubsection{Design of Trace Feature Representation}
The proposed trace feature representation comprises six channels, constructed by integrating three types of traffic statistical features. Each traffic statistic is further split into incoming and outgoing packet streams, resulting in a total of $3 \times 2 = 6$ feature channels. 
For ablation experiments, we isolate each type of traffic statistic (i.e., its corresponding incoming and outgoing channels) individually, along with the full fused feature set, yielding four distinct experimental configurations to quantify the contribution of each feature component.

\begin{figure}[ht]
    \centering
    \includegraphics[width=\columnwidth]{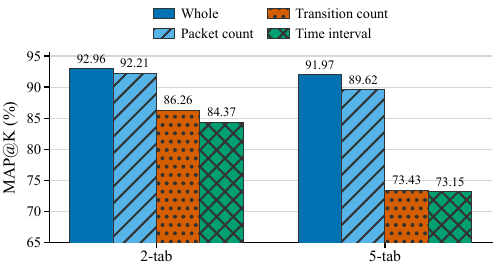}
    \caption{\added{Ablation of the robust trace representation in the closed-world 2-tab and 5-tab settings. Bars report mean MAP@$K$ for the complete representation and each individual feature group.}}
    \label{fig:feature_ablation}
\end{figure}

\added{As shown in Fig.~\ref{fig:feature_ablation}, the complete feature set performs best, with a mean MAP@2 of 92.96\% in the 2-tab setting and a mean MAP@5 of 91.97\% in the 5-tab setting. Among the individual feature groups, packet count performs best, with a MAP@2 of 92.21\% and a MAP@5 of 89.62\%. This result is consistent with prior WF studies that identify packet direction and count as highly discriminative traffic characteristics~\cite{sirinam2018deep, shen2023subverting}. Direction-transition count achieves a MAP@2 of 86.26\% and a MAP@5 of 73.43\%, while transition-time interval achieves a MAP@2 of 84.37\% and a MAP@5 of 73.15\%. The consistent advantage of the complete representation indicates that transition frequency and timing provide complementary information beyond packet counts, particularly under more complex 5-tab traffic interleaving.}

\begin{figure}[ht]
    \centering
    \subfloat[Loading time\label{fig:ablation_rep_a}]{
        \includegraphics[width=0.46\columnwidth]{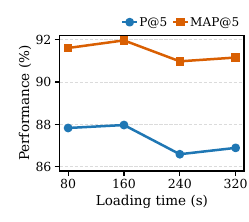}
    }
    \hfill
    \subfloat[Slot interval\label{fig:ablation_rep_b}]{
        \includegraphics[width=0.46\columnwidth]{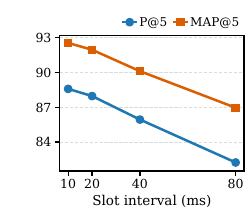}
    }
    \caption{\added{Sensitivity of PrismWF's robust trace representation in the closed-world 5-tab setting. (a) Performance versus maximum loading time at a fixed 20\,ms slot interval. (b) Performance versus slot interval at a fixed 160\,s loading window. The adopted configuration uses a 160\,s window with 20\,ms slots.}}
    \label{fig:ablation_feature_representation}
\end{figure}

\subsubsection{Impact of Maximum Loading Time}
We analyze the impact of the maximum loading time used in robust trace construction.
The maximum loading time determines how much traffic is retained from each trace and thus directly affects the completeness of temporal information available for WF attack.

\added{As shown in Fig.~\ref{fig:ablation_rep_a}, we evaluate maximum loading times of 80, 160, 240, and 320\,s in the closed-world 5-tab setting while fixing the slot interval at 20\,ms. At 80\,s, PrismWF achieves a P@5 of 87.83\% and a MAP@5 of 91.61\%. At 160\,s, P@5 and MAP@5 peak at 87.97\% and 91.97\%, respectively. At 240 and 320\,s, P@5 falls to 86.59\% and 86.89\%, respectively, while MAP@5 falls to 90.98\% and 91.16\%. These results suggest that 80\,s may omit useful late traffic, whereas longer windows provide no consistent gain and may introduce limited or noisy tail information. We therefore retain 160\,s.}

\subsubsection{Impact of Time Slot Interval}
\added{We next explore the effect of the time slot interval on trace discretization, a key parameter that controls the temporal resolution of the robust trace representation.}

\added{As shown in Fig.~\ref{fig:ablation_rep_b}, we evaluate slot intervals of 10, 20, 40, and 80\,ms under a fixed maximum loading time of 160\,s in the closed-world 5-tab setting. At 10\,ms, PrismWF achieves a mean P@5 of 88.59\% and a mean MAP@5 of 92.56\%. At 20\,ms, it achieves a mean P@5 of 87.97\% and a mean MAP@5 of 91.97\%. At 40 and 80\,ms, the mean P@5 decreases to 85.94\% and 82.23\%, respectively, while the mean MAP@5 decreases to 90.12\% and 86.96\%. This monotonic degradation is consistent with coarser temporal aggregation discarding discriminative timing details. We retain 20\,ms as the default interval, while acknowledging that 10\,ms achieves the highest mean performance.}

\subsubsection{Impact of Number of Blocks}
\begin{figure}[!t]
    \centering
    \includegraphics[width=0.96\columnwidth]{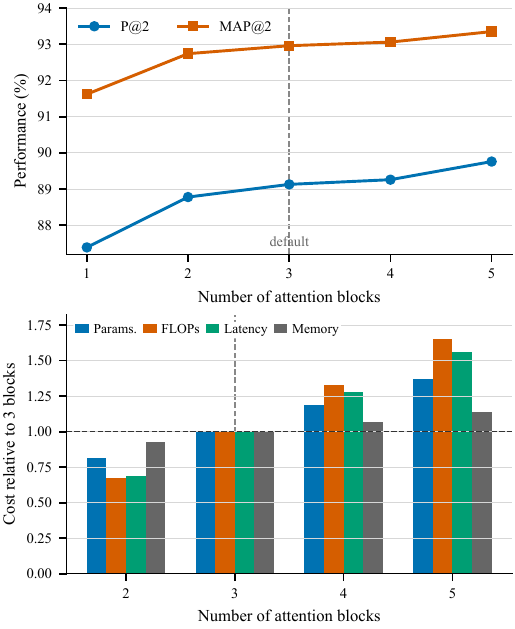}
    \caption{\added{Performance--cost tradeoff of PrismWF depth in the closed-world 2-tab setting. The upper panel reports the mean P@2 and MAP@2, and the lower panel normalizes four cost measures to the default three-block model.}}
    \label{fig:block_tradeoff}
\end{figure}

We investigate the influence of the number of stacked Multi-Granularity Attention Blocks on WF attack performance.
Specifically, we vary the number of blocks from 1 to 5 while keeping all other settings unchanged.
\added{As shown in Fig.~\ref{fig:block_tradeoff}, performance generally improves as more blocks are stacked, indicating that deeper multi-granularity interaction enables more effective refinement of traffic representations.}
\added{The upper panel reports the mean P@2 and MAP@2. With one block, PrismWF obtains a P@2 of 87.39\% and a MAP@2 of 91.62\%. Performance generally improves with depth, reaching a P@2 of 89.76\% and a MAP@2 of 93.35\% with five blocks. The default three-block model obtains a P@2 of 89.13\% and a MAP@2 of 92.96\%. These results are only 0.63 and 0.39 percentage points below those of the five-block variant, respectively, while using 40\% fewer attention blocks. Together with the cost measurements in the lower panel, this small performance gap supports three blocks as a balanced default configuration.}

\added{To explain the default depth, Fig.~\ref{fig:block_tradeoff} complements the performance results with cost measurements obtained under the same A800 GPU and inference protocol. The sweep was measured independently from Table~\ref{tab:efficiency}; its three-block latency of 29.19 ms and peak memory of 231.79 MB are consistent with the corresponding measurements in that table. Moving from two to three blocks improves mean P@2 and MAP@2 by 0.35 and 0.22 percentage points, respectively. Relative to the three-block model, a fourth block adds only 0.13 points in P@2 and 0.10 points in MAP@2, while increasing parameter count, profiled FLOPs, latency, and peak memory by 18.5\%, 32.8\%, 27.7\%, and 6.9\%. Five blocks achieve the highest mean P@2 and MAP@2, exceeding the three-block model by 0.63 and 0.39 percentage points, respectively, but require 37.1\% more parameters, 65.5\% more profiled FLOPs, 56.0\% higher latency, and 13.8\% more peak memory. We therefore adopt three blocks as the default setting because this configuration provides a favorable balance between performance and computational cost.}

\subsubsection{Sensitivity Analysis of Model Architecture}
\added{The core component of PrismWF is the \emph{Multi-Granularity Attention Block}, which combines fine-to-coarse interaction with router-level information exchange. We conduct controlled ablations on the original closed-world datasets from 2-tab to 5-tab. PrismWF w/o RI retains all four CNN branches and the fine-to-coarse interaction but disables router interaction. PrismWF w/o RI+GI further disables the fine-to-coarse interaction. PrismWF Single-G retains only the finest-grained CNN branch and its intra-granularity attention, and thus has no fine-to-coarse interaction. Except for the named component changes, all variants use the same robust trace representation, attention widths, patch--router mask, block depth, data partitions, and training protocol as the full model.}

\begin{table}[!t]
\centering
\caption{\added{Architecture ablation in the closed-world 2--5-tab settings.}}
\label{tab:arch_sensitivity}

\footnotesize
\setlength{\tabcolsep}{2.5pt}
\renewcommand{\arraystretch}{1.08}

{\color{black}
\resizebox{\columnwidth}{!}{
\begin{tabular}{cc|cccc}
\toprule
\textbf{Tabs} & \textbf{Metric}
& \textbf{w/o RI+GI} & \textbf{w/o RI}
& \textbf{Single-G} & \textbf{Full} \\
\midrule
\multirow{2}{*}{2} & P@2
& $86.50{\pm}0.56$ & $88.74{\pm}0.20$ & $\mathbf{89.31{\pm}0.13}$ & $89.13{\pm}0.38$ \\
& MAP@2
& $91.39{\pm}0.35$ & $92.70{\pm}0.12$ & $\mathbf{93.16{\pm}0.04}$ & $92.96{\pm}0.21$ \\
\midrule
\multirow{2}{*}{3} & P@3
& $81.19{\pm}0.84$ & $86.03{\pm}0.14$ & $86.01{\pm}0.17$ & $\mathbf{87.10{\pm}0.09}$ \\
& MAP@3
& $88.54{\pm}0.49$ & $91.42{\pm}0.14$ & $91.45{\pm}0.24$ & $\mathbf{91.71{\pm}0.20}$ \\
\midrule
\multirow{2}{*}{4} & P@4
& $82.37{\pm}0.35$ & $87.41{\pm}0.39$ & $87.30{\pm}0.18$ & $\mathbf{88.69{\pm}0.34}$ \\
& MAP@4
& $89.29{\pm}0.16$ & $92.17{\pm}0.30$ & $92.27{\pm}0.09$ & $\mathbf{92.60{\pm}0.28}$ \\
\midrule
\multirow{2}{*}{5} & P@5
& $78.60{\pm}0.81$ & $85.44{\pm}0.72$ & $85.26{\pm}0.29$ & $\mathbf{87.97{\pm}0.40}$ \\
& MAP@5
& $86.72{\pm}0.52$ & $90.79{\pm}0.44$ & $90.75{\pm}0.25$ & $\mathbf{91.97{\pm}0.37}$ \\
\bottomrule
\end{tabular}
}
}

\end{table}

\added{Table~\ref{tab:arch_sensitivity} shows that the contribution of the complete architecture becomes clearer as traffic-mixing complexity increases. In the 2-tab setting, the full model is slightly outperformed by Single-G, trailing it by only 0.18 percentage points in P@2 and 0.20 percentage points in MAP@2. From 3-tab onward, the full model achieves the highest mean P@$K$ and MAP@$K$ in every setting. As the tab count increases from two to five, router interaction raises P@$K$ over w/o RI by 0.39 to 2.53 points and raises MAP@$K$ by 0.26 to 1.18 points. Within the same four-branch backbone, the P@$K$ gain from fine-to-coarse interaction grows from 2.24 to 6.84 points, while the MAP@$K$ gain grows from 1.31 to 4.07 points. The variant without either interaction remains below Single-G, suggesting that parallel branches alone are insufficient without effective cross-granularity alignment. Overall, the widening gains suggest that coordinated multi-granularity extraction, fine-to-coarse aggregation, and router interaction are most useful under complex traffic mixing.}

\section{Discussion}
\label{sec6}

In this section, we discuss the limitations of the proposed method and outline promising directions for future work.

\noindent\textbf{Multi-Tab WF Attacks for Large-Scale Monitoring.}
In this work, we consider a moderate-scale setting with approximately 100 monitored websites.
Scaling multi-tab WF attacks to much larger monitored sets (e.g., hundreds or thousands of websites) remains a challenging problem.
As the number of target websites increases, traffic patterns become more diverse and label ambiguity becomes more severe, which may degrade classification performance.
A potential future direction is to incorporate structured relationships among websites (e.g., user browsing preferences or co-visitation patterns), which may further enhance scalability and robustness in large-scale multi-tab scenarios \cite{chen2019multi, ridnik2021asymmetric}.

\added{\noindent\textbf{Sensitivity to Dynamic Network Conditions.}
PrismWF adopts a fixed 20 ms slot interval to construct its traffic representation. Although this design provides a consistent temporal structure under the evaluated benchmark settings, substantial variations in network conditions may affect the resulting slot-level features. Severe packet loss may alter traffic statistics, while large timing variations may shift packet events across adjacent slots. In addition, the fixed observation window may omit delayed traffic under highly dynamic conditions. Evaluating PrismWF under live and time-varying Tor environments and exploring adaptive temporal representations remain important directions for future work.}

\added{\noindent\textbf{Evaluation Beyond Page-Load Scenarios.}
Our benchmark experiments and real-world Tor deployment primarily capture page loading and the subsequent network activity within a finite observation window. The deployment spans 96 contemporary websites and provides evidence under live Tor network conditions and contemporary web content. Long-lived interactive behaviors, including authenticated single-page application navigation, continuous scrolling, and prolonged media streaming, may exhibit different temporal characteristics and extend beyond the current observation window. Extending the evaluation to these continuously evolving traffic scenarios would further broaden the empirical scope of PrismWF.}

\added{\noindent\textbf{Multi-Tab Defenses and Real-World Deployment.}
Most existing WF defenses are primarily designed under the single-tab assumption \cite{juarez2016toward, holland2020regulator, gong2020zero, shen2024real}, whereas concurrent browsing introduces more complex traffic mixing and interleaving. 
We evaluate PrismWF through both benchmark experiments and real-world Tor deployments. Section~\ref{sec54} uses simulated defenses for comparability with prior multi-tab WF studies, whereas Section~\ref{sec:real_world} reports live 2-tab and 5-tab sessions under undefended traffic and online DeTorrent. The current deployment focuses on one online defense implementation, and future work will consider additional defenses and interactive browsing scenarios.}

\added{\noindent\textbf{Application-Layer Privacy Risks.}
PrismWF considers a passive network-layer adversary that infers visited websites from encrypted traffic without accessing browser state. Beyond this network-layer threat model, application-layer browser vulnerabilities can independently weaken user unlinkability. For example, CVE-2026-6770 exposed a cross-origin correlation issue in affected Firefox and Tor Browser versions, underscoring that Tor users may face privacy risks at both the network and application layers.}

\section{Conclusion}

\added{In this paper, we propose PrismWF, a high-performance multi-tab WF attack method tailored for traffic mixing scenarios in multi-tab browsing.
Specifically, PrismWF first constructs a robust traffic feature representation, then leverages a multi-branch CNN to extract traffic features at different granularities, and finally employs a multi-granularity attention block designed to capture the characteristics of mixed multi-tab traffic, thereby enabling effective website identification.
Extensive experiments demonstrate that PrismWF achieves state-of-the-art performance across closed-world, open-world, mixed-tab, unknown-$K$, and representative simulated-defense settings, with stable gains over repeated training runs. We further evaluate PrismWF in a real-world Tor deployment, where it achieves the highest P@$K$ in all four conditions and the highest MAP@$K$ in three.}
For future work, we plan to extend the proposed method to support large-scale monitored website sets and to explore dedicated defense mechanisms targeting multi-tab WF attacks.

\label{sec_conclusion}

\IEEEtriggercmd{\enlargethispage{2\baselineskip}}
\IEEEtriggeratref{43}
\bibliography{ref}
\bibliographystyle{IEEEtran}

\ifCLASSOPTIONcaptionsoff
  \newpage
\fi

\end{document}